\documentclass[aps,twocolumn,showpacs]{revtex4}
\usepackage{amssymb,amsmath,bm,enumerate,graphicx,color,subfigure}
\begin{document}
\title{L\'{e}vy flights in confining environments: Random paths  and their statistics}
\author{Mariusz  \.{Z}aba,  Piotr Garbaczewski  and  Vladimir Stephanovich}
\affiliation{Institute of Physics, University of Opole, 45-052 Opole, Poland}
\date{\today }
\begin{abstract}
We analyze  a specific class of   random  systems   that  are  driven  by  a symmetric  L\'{e}vy stable noise. In view of   the  L\'{e}vy
noise sensitivity to the    confining   "potential landscape"  where jumps take place (in other words, to environmental inhomogeneities),
 the pertinent  random motion asymptotically sets down at  the  Boltzmann-type equilibrium, represented by a  probability density function (pdf)  $\rho _*(x) \sim \exp [-\Phi (x)]$. Since there is no Langevin representation of the dynamics in question,
  our main goal here is to establish   the   appropriate   path-wise description  of the underlying jump-type process and  next
   infer the  $\rho (x,t)$  dynamics directly from the random paths statistics.  A priori  given data are jump transition rates entering
   the master equation  for  $\rho (x,t)$  and its target  pdf  $\rho _*(x)$.  We use numerical methods and  construct
    a   suitable modification of the Gillespie algorithm,  originally invented  in  the  chemical kinetics context.
The   generated   sample trajectories show up a qualitative typicality, e.g.  they  display structural   features  of
  jumping paths (predominance of small vs large  jumps) specific to  particular    stability indices  $\mu \in (0,2)$.
\end{abstract}
\pacs{05.40.Jc, 02.50.Ey, 05.20.-y, 05.10.Gg}
\maketitle

\section{Introduction}
Many  random processes  in real physical systems  admit  a  simplified   description based on stochastic differential equations.
In such case there is a routine  passage  procedure  from microscopic   random variables to macroscopic (statistical  ensemble) data.
 The latter are   encoded in the   time evolution of an associated probability density function (pdf)  which  is  a solution of  a deterministic  transport equation.
A paradigm example  is  the  so-called Langevin modeling of diffusion-type and jump-type processes.
The presumed microscopic model of the dynamics in external force fields  is provided by the Langevin (stochastic)  equation whose
direct  consequence is   the  Fokker-Planck   equation, \cite{risken} and \cite{fogedby}.
 We note that in case of jump-type processes the familiar  Laplacian (Wiener noise generator) needs
  to be replaced by a suitable pseudo-differential operator (fractional Laplacian, in case of a symmetric   L\'{e}vy-stable noise).

 We pay a particular attention to jump-type processes which are omnipresent  in Nature (see \cite{metzklaf} and references therein).  Their characterization is  primarily provided by jump  transition rates between different states of the system under consideration.
   However our major  focus is  on a  specific  class of random systems which are   plainly incompatible with   a straightforward Langevin modeling of jump-type processes and,  as such,  are seldom  addressed in the literature.

To this end we depart  from the  concept, coined in an  isolated  publication  \cite{deem},  of
L\'{e}vy flights-driven models of disorder that, while at equilibrium, do  obey detailed balance.
The corresponding  research  line  has been effectively  initiated in Refs.  \cite{brockmann}-\cite{belik}.
It has  next  been expanded in various directions,   with
a special  emphasis  on so-called L\'{e}vy-Schr\"{o}dinger  semigroup reformulation of  the original    probability density function
(pdf)  dynamics,  \cite{vilela1,vilela2,geisel}  and
 \cite{olk}-\cite{gar},  c.f. also \cite{lorinczi,vilela1,vilela2}.  We note in passing that the familiar
  Fokker-Planck  equation can be equally well formulated in terms of the  Schr\"{o}dinger semigroup and this property is universally valid in
 the  standard theory of Brownian motion,  \cite{risken,lorinczi}. Its generalization to L\'{e}vy   flights is neither  immediate nor obvious.
  It is   often considered in the  prohibitive vein following \cite{klafter,cohen}.

In fact, in relation to  L\'{e}vy flights,  a  novel   fractional  generalization  of the   Fokker-Planck equation has been introduced in Refs. \cite{brockmann}-\cite{belik}
    to handle systems that are randomized  by symmetric  L\'{e}vy-stable  drivers. In this case,  contrary to  the  popular lore  about properties of
     (Langevin-based) L\'{e}vy processes c.f. Refs. \cite{klafter}-\cite{dubkov} and \cite{cohen},  the pertinent   random systems  are allowed to
      relax to   (thermal) equilibrium states of a standard Boltzmann-Gibbs form.

      The  underlying  jump-type processes, in the stationary (equilibrium)  regime,
 respect the  principle of  detailed balance  by construction \cite{gar}.    Their distinctive feature,  if compared with  the standard Langevin modeling
 of L\'{e}vy flights,
is that  they have a built-in response  {\it not}  to external forces {\it  but rather to} external force potentials.
These potentials are interpreted  to form  confining "potential landscapes" that are specific to the environment.  L\'{e}vy  jump-type
 processes  appear to be particularly sensitive to environmental inhomogeneities, \cite{brockmann,gar2}.

L\'{e}vy  flights  are pure jump (jump-type) processes. Therefore, it seems useful  to indicate  that various model realizations of  standard   jump processes (jump size is bounded from below and above) can be thermalized  by means of a specific scenario of an energy exchange with the thermostat. It is
based on the {\it  principle of detailed balance}.
We have discussed this issue in some detail before \cite{gar} along with an extension of  this conceptual framework  to L\'{e}vy-stable processes. Not to reproduce   easily   available   arguments of past publications, we shall be very rudimentary in our motivations.

We quantify a  probability density evolution, compatible with a  jump-type process on $R$ (this limitation may in principle be lifted in favor of $R^n$), in terms of the master  equation:
\begin{equation}
\partial_t\rho(x)=\int\limits_{\varepsilon_1\leq |x-y|\leq \varepsilon_2} [w_\phi(x|y)\rho(y)-w_\phi(y|x)\rho(x)]dy,\label{l1}
\end{equation}
where $\varepsilon_1$ and $\varepsilon_2$ are, respectively, the lower and upper bounds of jump size and
\begin{eqnarray}
w_\phi(x|y)&=&C_\mu\frac{\exp[(\Phi(y)-\Phi(x))/2]}{|x-y|^{1+\mu}},\nonumber \\ \nonumber \\
C_\mu&=&\frac{\Gamma(1+\mu)\sin(\pi\mu/2)}{\pi} \label{l2}
\end{eqnarray}
is the  jump   transition rate  from  $y$  to  $x$.  We stress that  $w_\phi(x|y) $ is a  non-symmetric function of $x$ and $y$.

An implicit Boltzmann-type weighting involves a square root of a target pdf  $\rho _*(x) \sim  \exp[-\Phi  (x)]$ and  accounts for
the  a priori  prescribed   "potential landscape" $\Phi  (x)$  whose confining features affect the  jump-type process.
What  matters is  a relative  impact   of a confinement strength   of $\Phi (x)$  (level of attraction, see Ref. \cite{belik})
 upon jumps of  the size  $|x-y|$,    both at  the   point  of origin  $y$ and that of  destination   $x$.
  In principle, $\Phi(x)$ may be an arbitrary function that secures a  $L^1(R)$  normalization of $\exp(-\Phi(x))$.
  In this case, the resultant pdf $\rho _*$  is a  stationary solution of the transport equation
   (\ref{l1}) with  unbounded jump length, e.g.  $\varepsilon_1 \to 0$   and  $\varepsilon_2 \to\infty$.

 We note that the presence of lower and upper bounds of the jump size $\varepsilon_{1,2}$, that are necessary for an implementation
 of  numerical algorithms, enforces a truncation of  the jump-type  process  (without any cutoffs) to a standard  jump process. The transition rates of the latter,  however, are ruled by
 L\'{e}vy measures of symmetric  L\'{e}vy stable noises with  $\mu \in (0,2)$.
 A lower bound for  the jump size  is usually removed while evaluating the corresponding integrals in the  sense of their Cauchy principal values.
An upper bound is less innocent and its effects need to be controlled by long tailed  pdfs which  stands for a distinctive
feature of L\'{e}vy flights, see a discussion of L\'{e}vy stable limits of step  processes in Ref.  \cite{olk}.
There is also pertinent discussion of a long time behavior of   (unconfined, e.g. free)  truncated L\'{e}vy flights in Ref. \cite{mantegna}.

In contrast to procedures based on the Langevin modeling of L\'{e}vy flights in external force fields, \cite{fogedby,chechkin,dubkov},
 there is no known path-wise approach underlying  the transport equation  (\ref{l1}). With  no  direct  access to  sample  trajectories
  of the stochastic process in question,  a  method must be devised to generate random paths directly
   from jump transition rates (\ref{l2}).  The additional requirement here is that we set
    a priori a  "potential landscape" $\Phi (x)$ for a  chosen  jump-type (symmetric L\'{e}vy stable) noise driver.

The outline of the paper is as follows.   First we describe our modification of the Gillespie algorithm which entails  a numerical generation
of random paths for the dynamics determined by Eqs  \eqref{l1} and \eqref{l2}.  Next the statistics of random paths is addressed and
various accumulated data are analyzed with a focus on  inherent compatibility issues.

We  analyze  generic  (Cauchy, quadratic Cauchy)  and non-generic  (Gaussian  and locally periodic)  examples of target pdfs  for the jumping dynamics.
 Random paths are generated in conjunction with  representative L\'{e}vy stable drivers, like e.g. those  indexed by  $\mu =1/2, 1,  3/2$.
 Their qualitative typicality is  emphasized.

Statistical data, acquired from our modification of Gillespie algorithm, have been  employed to generate the dynamical patterns
of  behavior $\rho (x,t) \to \rho _*(x)$,  to demonstrate the compatibility of the transport (master)  equation
 \eqref{l1}, \eqref{l2}  and  its  underlying   path-wise representation.
 Both coming from  the  predefined  knowledge of the target pdf and non-symmetric (biased)  jump transition rates.

\section{Random paths: Modified Gillespie algorithm.}

Here we adopt   \cite{sokolov} (and  properly adjust to handle L\'{e}vy  flights) basic tenets of  so-called
Gillespie's algorithm \cite{gillespie,gillespie1}.
Originally, this algorithm had been devised to simulate  random properties of coupled chemical reactions.
The advantage of the algorithm is that it permits to generate  random trajectories of the  corresponding stochastic process
directly from its (jump) transition rates,  with  no need for any  stochastic differential equation and/or its explicit solution.
 We emphasize that this  feature  of Gillespie's algorithm is vitally important, since Langevin modeling is not  operational
  in our framework.

We   rewrite Eq. (\ref{l1})  in  the form ($x-y=z$)
\begin{eqnarray}
&&\partial_t\rho(x)=\int\limits_{\varepsilon_1\leq|z|\leq \varepsilon_2} \biggl[w_\phi(x|z+x)\rho(z+x)-\nonumber \\
&&-w_\phi(z+x|x)\rho(x)\biggr]dz.\label{l3}
\end{eqnarray}
To construct a reliable path generating algorithm consistent with Eq. (\ref{l3}) we first note that
chemical reaction channels in the original Gillespie's  algorithm  may be re-interpreted as jumps  from one spatial point
 to another, like transition channels in the spatial  jump process. An obvious provision is that the set of possible chemical
 reaction channels is finite (and generically low), while we are interested in all admissible jumps from a chosen point
 of origin  $x_0$ to any of
  $[x_0-\varepsilon_2,x_0-\varepsilon_1]\cup[x_0+\varepsilon_1,x_0+\varepsilon_2]$.  It is clear  that such jumps form an infinite  continuous set.
With a  genuine computer simulation in mind, we must respect   standard   numerical assistance  limitations.
Surely we cannot  admit all conceivable jump sizes. As well, the number of destination points, even if potentially enormous,
must remain finite for any fixed point of origin.

Our modified version of the  Gillespie's algorithm, appropriate for handling of spatial  jumps is as follows \cite{available}:
\begin{enumerate} [(i)]
\item Set time  $t=0$ and the point of origin  $x=x_0$.
\item Create the set of all admissible  jumps from $x_0$ to $x_0+z$ that is compatible with the transition rate  $w_\phi(z+x_0|x_0)$. \\
\item Evaluate
\begin{eqnarray}
&&W_1(x_0)=\int_{-\varepsilon_2}^{-\varepsilon_1}w_\phi(z+x_0|x_0)dz, \nonumber \\
&&W_2(x_0)=\int_{\varepsilon_1}^{\varepsilon_2}w_\phi(z+x_0|x_0)dz \label{l4}
\end{eqnarray}
and  $W(x_0)=W_1(x_0)+W_2(x_0)$.
\item Using  a random number generator draw  $p\in[0,1]$  from a uniform distribution.
\item Using above $p$   and identities
\begin{widetext}
\begin{equation}
\left\{
  \begin{array}{ll}
    \int\limits_{-\varepsilon_2}^{b}w_\phi(z+x_0|x_0)dz=p W(x_0), & \hbox{$p<W_1(x_0)/W(x_0)$;} \\
    W_1(x_0)+\int\limits_{\varepsilon_1}^{b}w_\phi(z+x_0|x_0)dz=p W(x_0), & \hbox{$p\geqslant W_1(x_0)/W(x_0)$,}\label{l5}
  \end{array}
\right.
\end{equation}
\end{widetext}
find  $b$ corresponding to the  "transition channel"  $x_0 \rightarrow  b$.
\item  Draw a new number  $q\in(0,1)$  from a uniform distribution.
\item Reset time label  $t=t+\Delta t$  where  $\Delta t=-\ln q/W(x_0)$.
\item Reset $x_0$ to a new value $x_0+b$.
\item Return  to step (ii) and repeat the procedure  anew.
\end{enumerate}

\textbf{Comment 1}: The original Gillespie algorithm employs a discrete label  $\nu$ (with a finite range)
 enumerating possible chemical reactions channels. To identify a channel, one must look  for estimates of a double inequality
  (see Eq. (21b)  of Ref. \cite{gillespie1})
\begin{equation} \label{summ}
\sum\limits_{\nu=1}^{\mu-1}a_\nu<r_2a_0\leqslant \sum\limits_{\nu=1}^{\mu}a_\nu,\qquad a_0=\sum\limits_{\nu=1}^M a_{\nu},
\end{equation}
where $r_2$    is a random number,  $M$  indicates a total number of chemical reaction channels
and  $a_\nu dt$  stands for a probability  that  the  $\nu$-th reaction  would actually take place in the interval $(t,t+dt)$.
To adjust this recipe to our settings, we  need to enumerate the infinite number of (infinitely close) channels.
This corresponds to passing from summation to integration in Eq. \eqref{summ}. As it has been pointed out above,
such situation corresponds to possible transitions from $x_0$   into an interval
$[x_0-\varepsilon_2,x_0-\varepsilon_1]\cup[x_0+\varepsilon_1,x_0+\varepsilon_2]$.
As the Lebesgue measure of a point equals zero, we can replace inequalities by identities in \eqref{summ},
see step  (v)  of the above algorithm.   Formally, we can say  that although the number of jumps destinations is finite,
 their number is so large that  it is consistent  to approximate it by a dense subset
  of intervals  $[x_0-\varepsilon_2,x_0-\varepsilon_1]\cup[x_0+\varepsilon_1,x_0+\varepsilon_2]$.
  This justifies a replacement of finite sum by an integral over corresponding interval.
 The following jump size bounds  (integration boundaries) were adopted in the numerical procedure:
 $\varepsilon_1=0.001$ and  $\varepsilon_2=1$.

\section{Statistics of random  paths: pdf time evolution   and   compatibility   issues.}
Our  main  task in the present section is to select,  to some extend generic,   transition rates
for jump-type  processes that will prove to be amenable to the outlined  random path generation procedure.
Once suitable path ensemble data  are collected,    we shall  verify whether
statistical  (ensemble)   features of generated random trajectories  are  compatible with the master equation  (\ref{l1}).
 That includes a control  of  an asymptotic behavior   $\rho (x,t) \to \rho_*(x)$ when  $t\to\infty$.

\subsection{Harmonic confinement (Gaussian target)}
Let us consider an asymptotic invariant (target)  pdf in the Gaussian form:
\begin{equation}
\rho_*(x)=\frac{1}{\sqrt{\pi}}e^{-x^2}.\label{l6}
\end{equation}
The corresponding $\mu $-family of transition rates reads
\begin{equation}
w_\phi(z+x|x)=C_\mu\frac{e^{-z^2/2+x z}}{|z|^{1+\mu}}.\label{l7}
\end{equation}
According to the step (iii) of the simulation algorithm, we must   evaluate integrals with transition rates  (\ref{l7}) in the intervals $[-\varepsilon_2,-\varepsilon_1]$  and
 $[\varepsilon_1,\varepsilon_2]$.
To execute the  step (iv) of the algorithm, we employ the Mersenne-Twister random number generator, \cite{mersenne}.
We find  $b$ by a numerical solution of the transcendental equation \eqref{l5} in conformity with step (v) of the algorithm.
 The C-codes for trajectory generating algorithm, \cite{available},   were finally employed to get  the  trajectory statistics
 data for three specific choices of L\'{e}vy drivers, namely
$\mu=0.5, 1, 1.5$.

\begin{figure*}
\centerline{
\includegraphics[width=0.7\columnwidth]{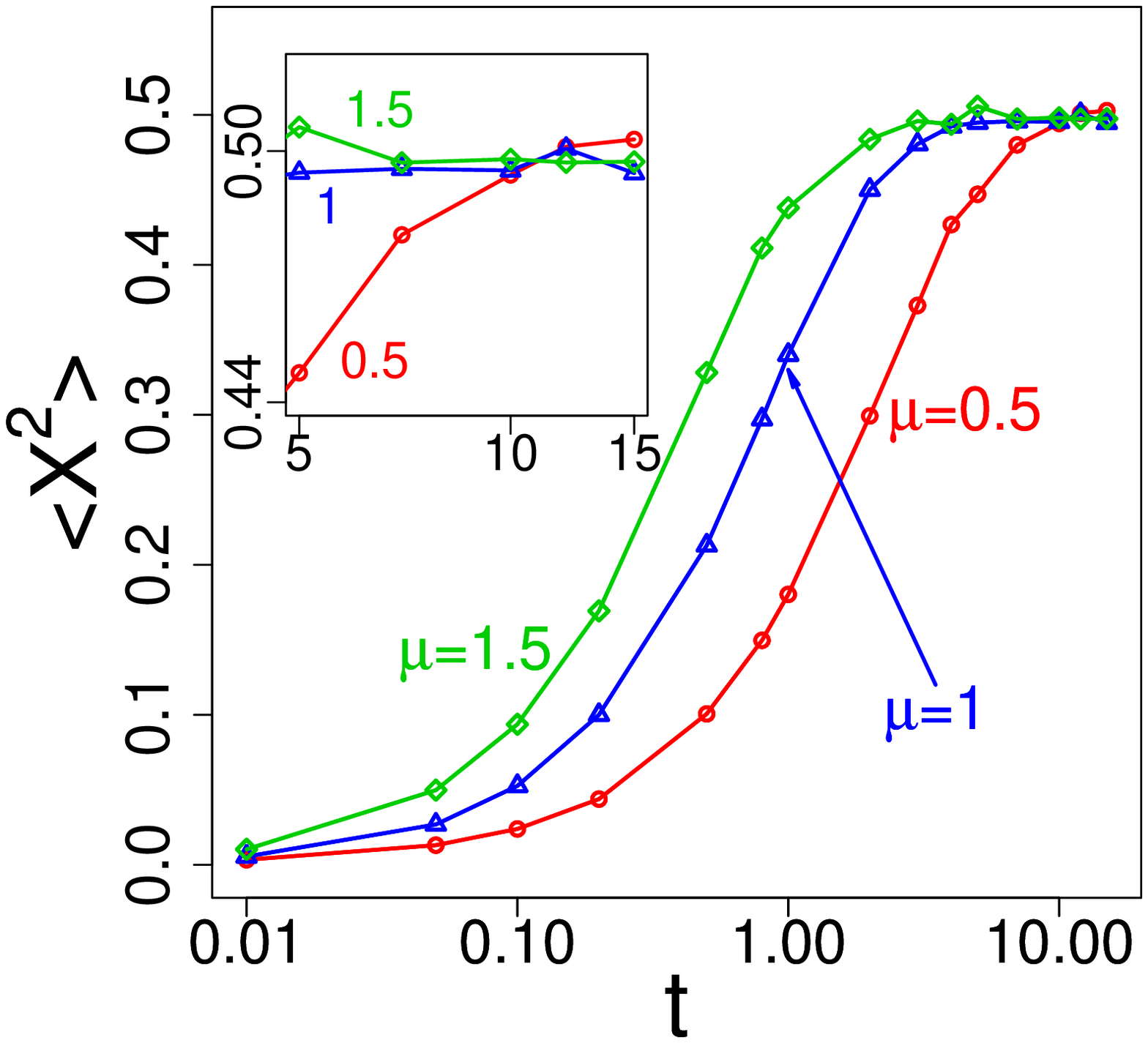}
\includegraphics[width=0.7\columnwidth]{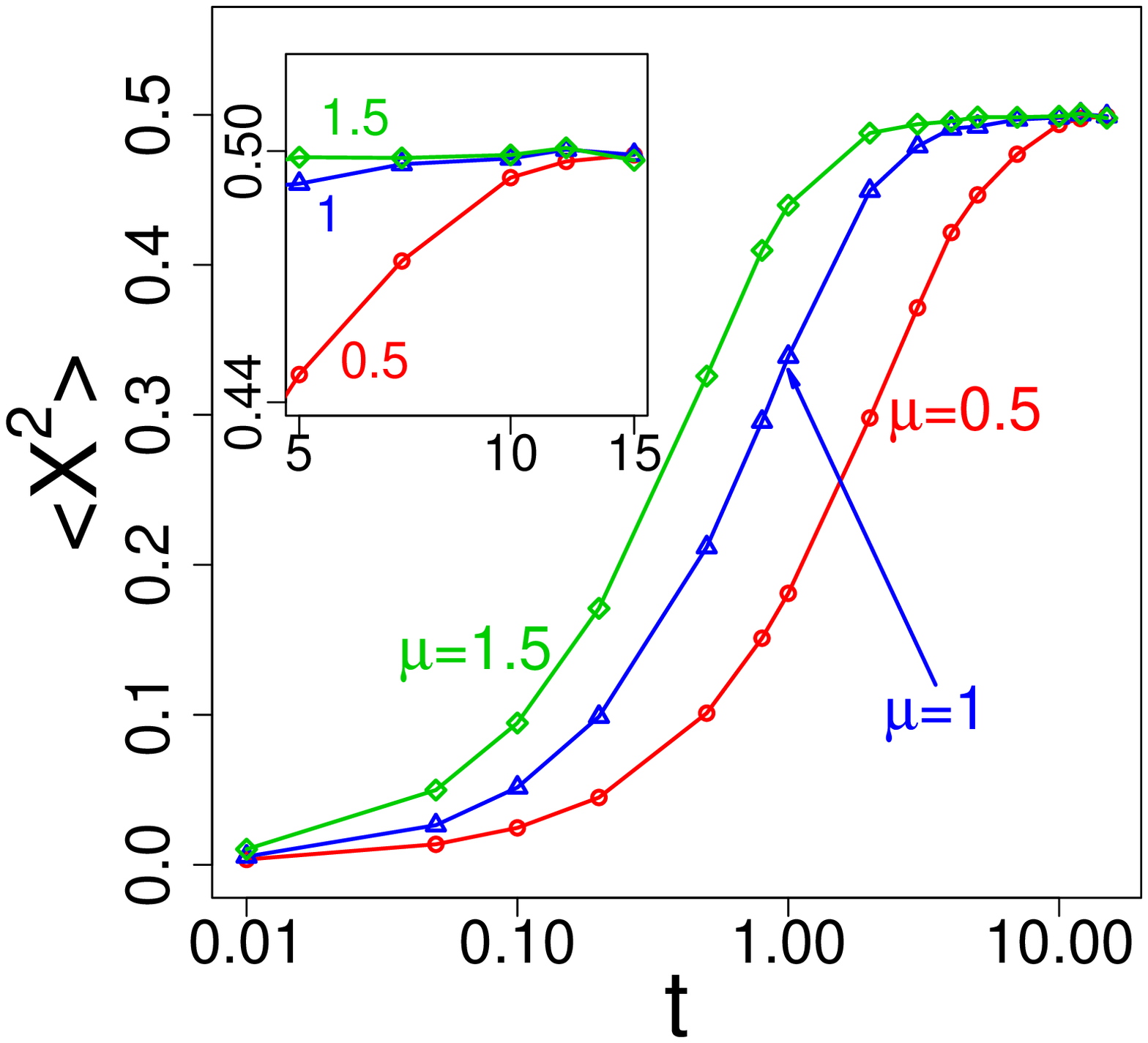}
\includegraphics[width=0.7\columnwidth]{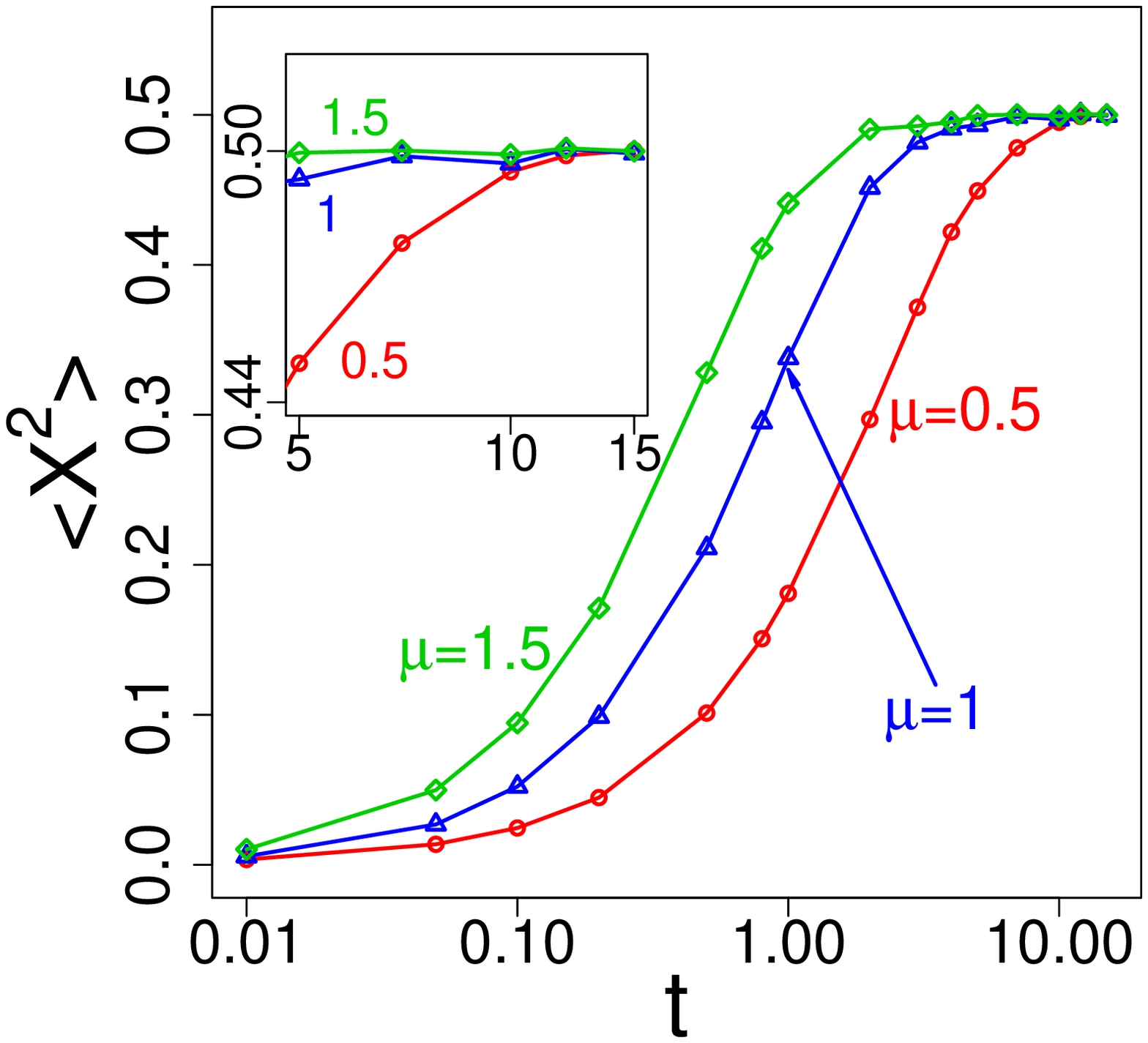}}
\caption{Gaussian target: Time evolution of  the pdf   $\rho (x,t)$  second moment   for  25 000 (left panel), 50 000 (middle panel) and 75 000 (right panel) trajectories. Insets visualize the oscillations smoothing in the asymptotic regime for $10\leq t\leq 15$; figures near curves correspond to $\mu$ values.}
\label{rys}
\end{figure*}

\textbf{Comment 2}:  For small $z$, transition rates vary rapidly so that adaptive numerical integration algorithms become very time-consuming.
 To speed up the calculations, we propose a more efficient  procedure (than adaptive numerical integration algorithms with huge number of subdivisions
 for small $z$), that amounts to separate integrations for  "large" and  "small"  $z$ subintervals. At small $z$ we can expand the
  corresponding integrand in Taylor series and truncate it at, say, quadratic term. Generally speaking, the number of terms to
   be left depends on the accuracy which should be retained during the integration.
    At "large" $z$ the transition rates vary gradually so that standard adaptive numerical integration  works fine.

 As an example, we consider  the  $[\varepsilon_1,\varepsilon_2]$  integration with  $\mu=1$.  We set  $\varepsilon_{12}=0.05$,
  so that for   $z \in [\varepsilon_1,\varepsilon_{12}]$  we get
\begin{eqnarray}
&&\int\limits_{\varepsilon_1}^{\varepsilon_{12}}\frac{e^{-z^2/2-x z}}{z^{2}}dz\thickapprox
\int\limits_{\varepsilon_1}^{\varepsilon_{12}}\frac{1-xz+\frac{x^2-1}{2}\,z^2}{z^2}dz=\nonumber \\
&&=\frac{\varepsilon_{12}-\varepsilon_1}{2\varepsilon_1\varepsilon_{12}}\left[2+(x^2-1)\varepsilon_1\varepsilon_{12}\right]-x\ln\left|\frac{\varepsilon_{12}}{\varepsilon_1}\right|.
\label{l7a}
\end{eqnarray}
In the interval   $z\in[\varepsilon_{12},\varepsilon_2]$ we evaluate  the integral numerically.
The proposed hybrid   procedure  (integrating analytically in the "most dangerous" small $z$ interval and numerically otherwise)
 permits to speed up the calculation drastically and has actually been used  in our simulations.

\begin{figure*}
\centerline{
\includegraphics[width=0.7\columnwidth]{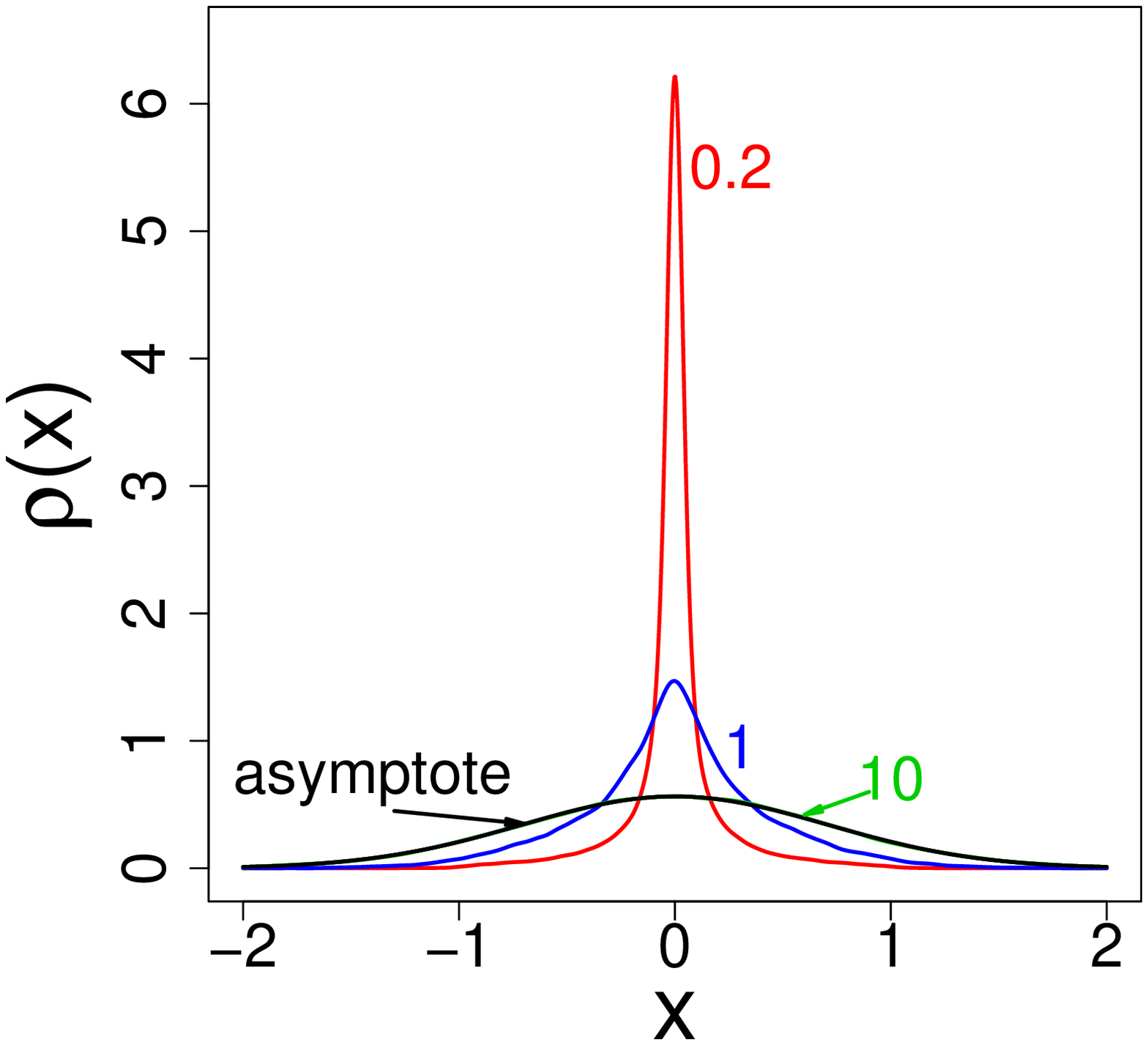}
\includegraphics[width=0.7\columnwidth]{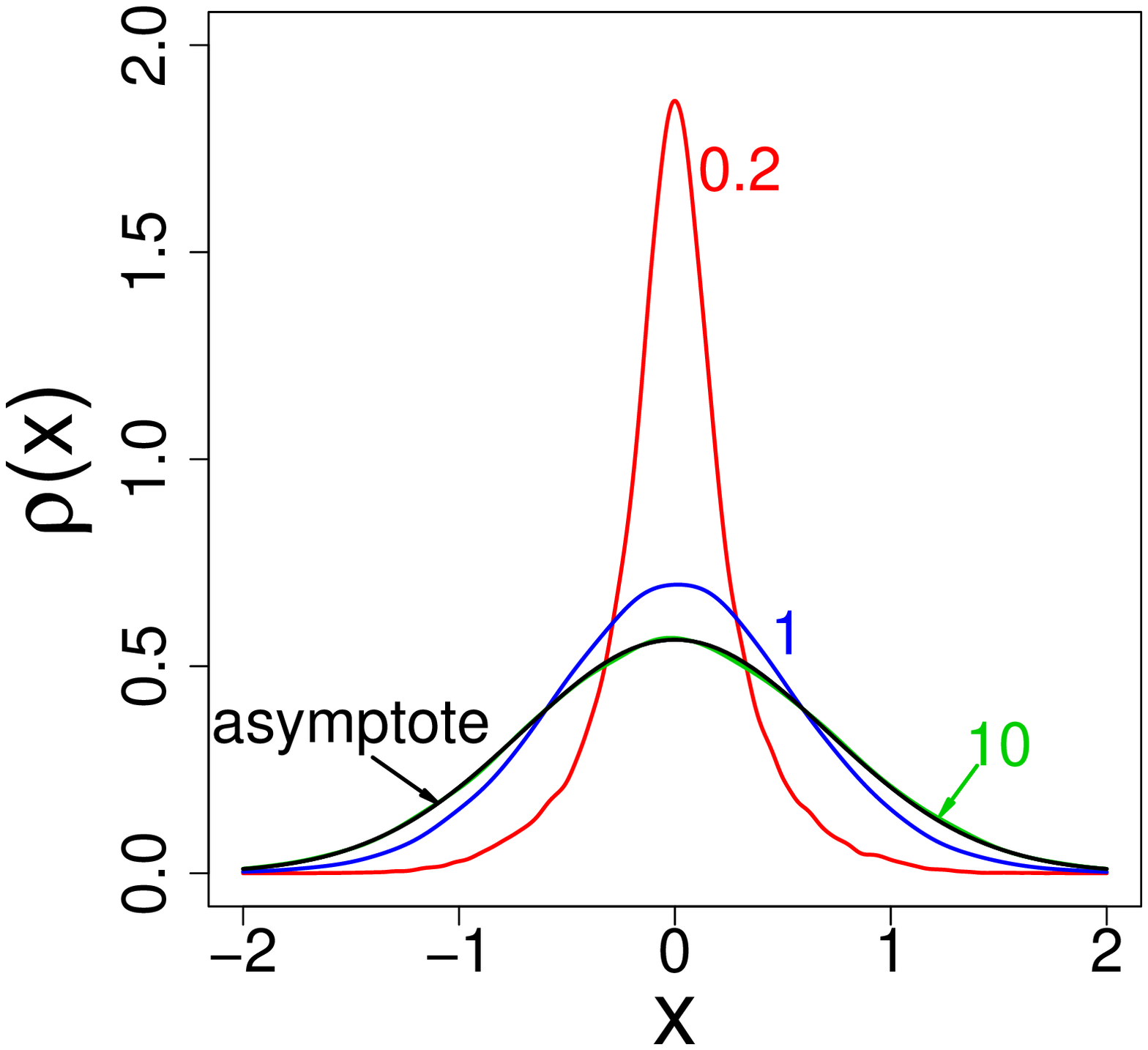}
\includegraphics[width=0.7\columnwidth]{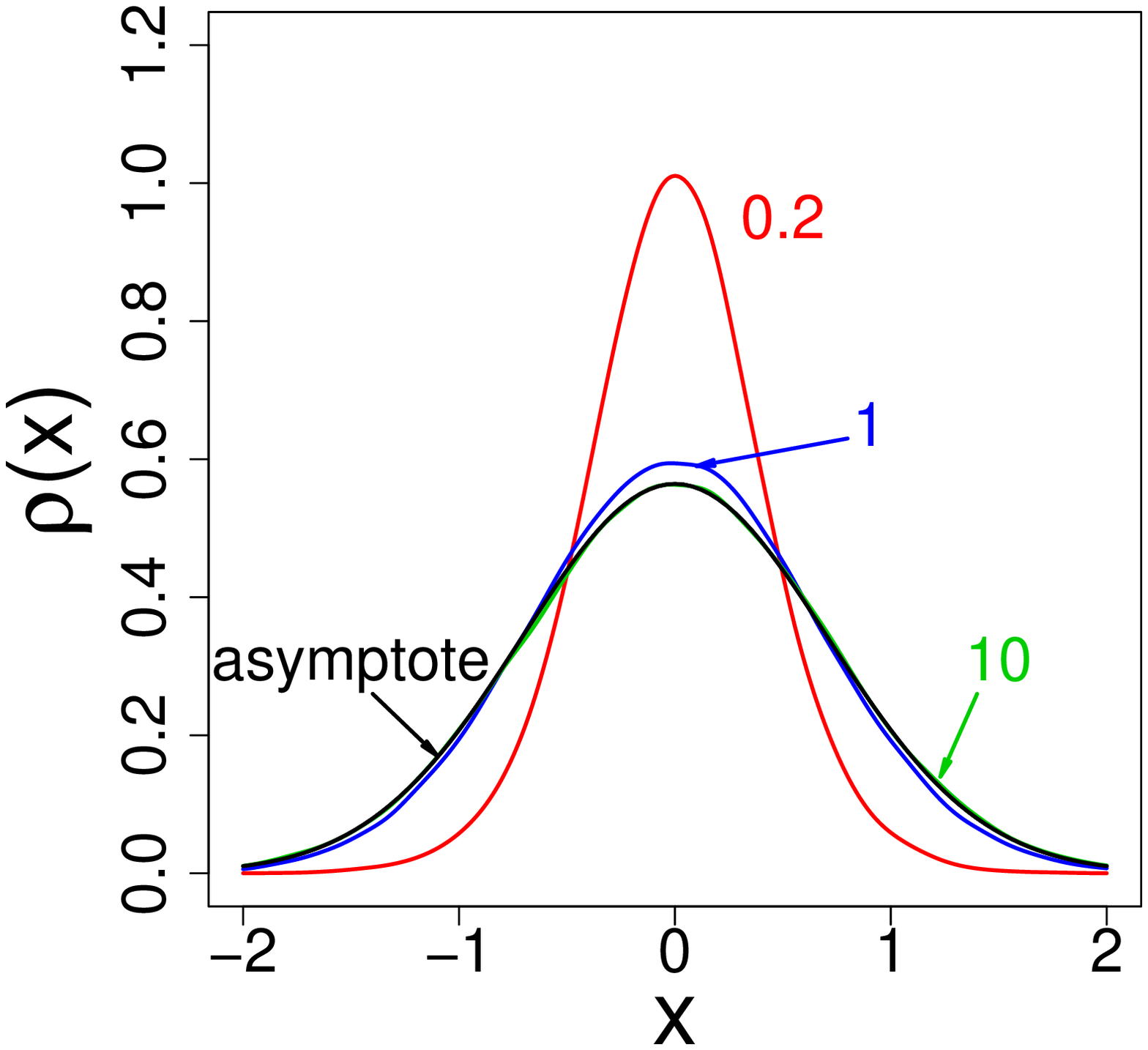}}
\caption{Gaussian target: Time  evolution of $\rho (x,t)$  inferred from  75 000 trajectories:  $\mu=0.5$ (left panel), $\mu=1$ (middle panel) and $\mu=1.5$ (right panel).  All trajectories originate from $x=0$,  i.e. refer to the $\delta(x)$-type initial probability distribution.}
\label{rys2}
\end{figure*}

\begin{figure*}
\centerline{
\includegraphics[width=0.7\columnwidth]{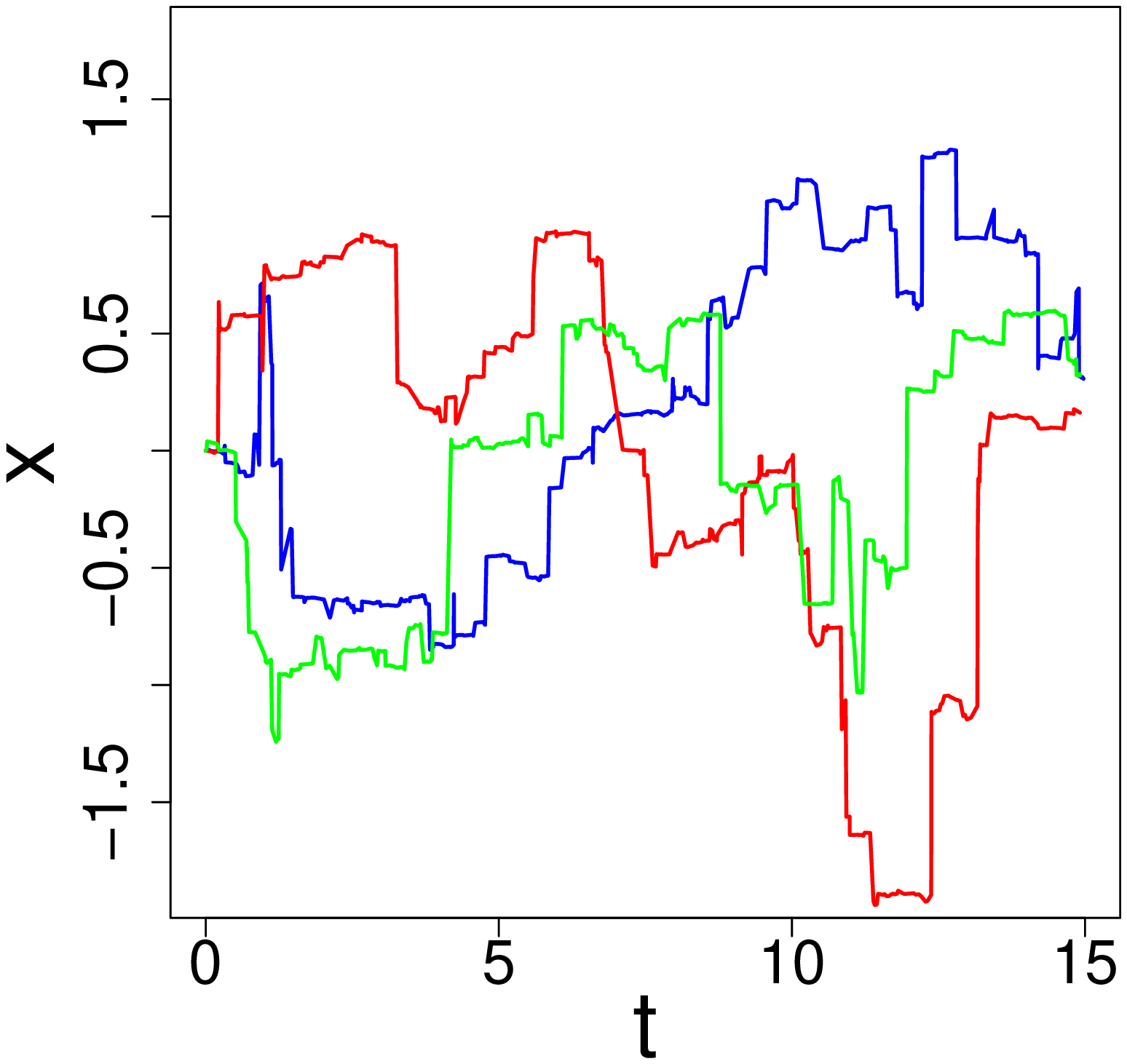}
\includegraphics[width=0.7\columnwidth]{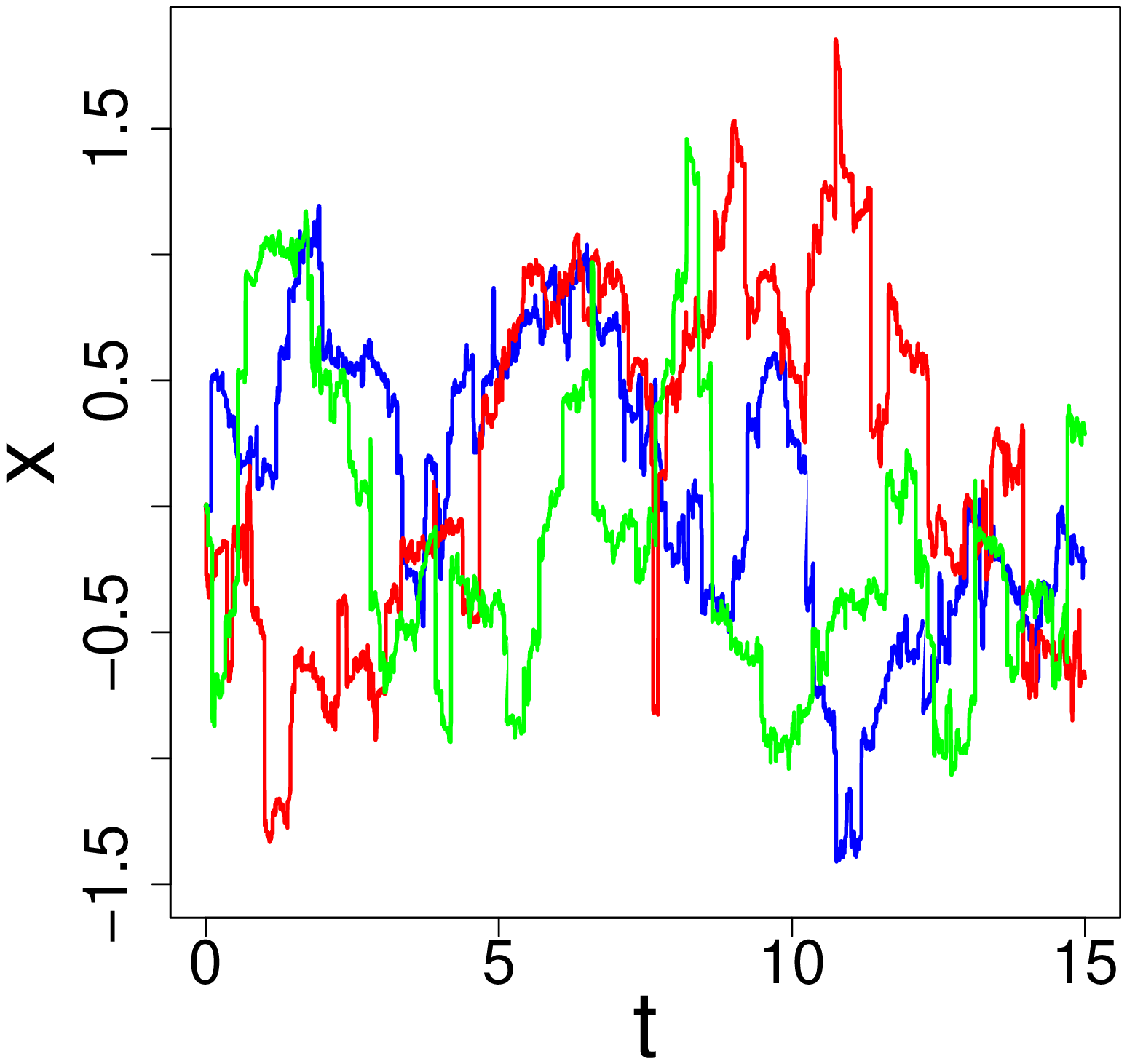}
\includegraphics[width=0.7\columnwidth]{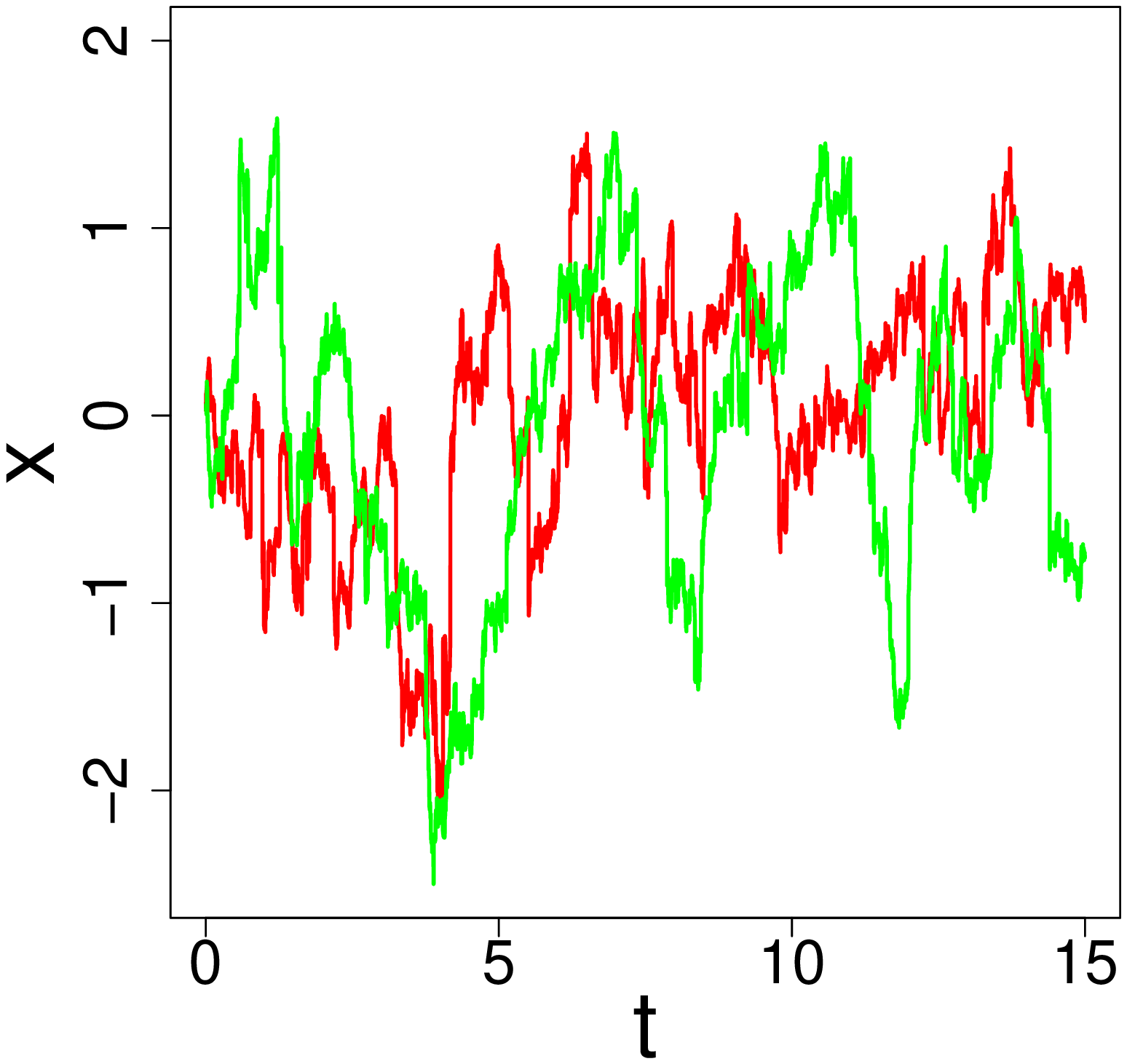}}
\caption{Gaussian target. Qualitative typicalities of sample paths  for $\mu=0.5$ (left panel), $\mu=1$ (middle panel) and $\mu=1.5$ (right panel).
 All trajectories originate form $x=0$. Right panel comprises two trajectories instead of three (for better visibility).}
\label{rys3}
\end{figure*}

The results of our numerical simulations are reported  in Figs. 1 through  3.
We note, that on Fig. 1 the second moment oscillates near its equilibrium value $1/2$. The oscillations are smoothed out with the growth of
 the number of random   trajectories that contribute to the statistics. A numerical convergence to $<X^2>=$$1/2$ is consistent
  with an analytic equilibrium value of the second moment  of  the chosen $\rho_*(x)$ \eqref{l6}. The rate of this convergence is
   higher for larger $\mu \in (0,2)$.
Clearly, for small $\mu$ the big jumps are frequent which  enlarges  the inferred time intervals  $\Delta t$
 in the Gillespie's algorithm, see the trajectories on left panel of Fig. 3. Thus,  the relaxation  to equilibrium is slow.
 It gets   faster for  larger $\mu $, when big jumps are rare and time intervals  $\Delta t$  are  generically very small.

Fig. 2 displays a probability density evolution, inferred from the ensemble  statistics of   $75 000$  trajectories.
 All of them have started form the same point $x=0$.  Although the data  fidelity grows with  the number of contributing paths,
 we have not found significant  qualitative differences to justify a presentation of  data for 100 000, 200 000, 250 000 and more trajectories.
  The relaxation  time rate dependence on  $\mu $ is clearly visible as well. It suffices to analyze  differences between three curves for $t=0.2$ and/or  $t=1$. We observe a conspicuous lowering of their maxima  with  the growth
  of $\mu $ (take care of different scales on the vertical axes on Fig.2 panels).
The simulated  pdfs at $t=10$  are  practically indistinguishable from an exact analytical  asymptotic pdf \eqref{l6}.
The  convergence of $\rho (x,t)$   towards    $\rho_*(x)$  appears  to be  relatively fast irrespective of the chosen $\mu $-driver.

Although our reasoning is definitely path-wise and all data have been extracted from trajectory ensembles,
 it is instructive to visualize generic sample paths. That is accomplished in Fig. 3, basically to indicate their (paths)
 qualitative typicalities. The structural impact of larger against smaller jumps  can be visually compared   and has been
  found to conform with  standard simulations of L\'{e}vy stable sample paths (with no forces or potentials involved), c.f. \cite{janicki}.

\subsection{Logarithmic confinement}
\subsubsection{Quadratic Cauchy target}

Let us consider  a long-tailed asymptotic pdf which is a special  $\alpha =2$ case of the  one-parameter
$\alpha $-family of equilibrium (Boltzmann-type) states,  associated with a logarithmic potential  $\Phi (x) \equiv  \alpha \ln (1+x^2)$, $\alpha  > 1/2$, see \cite{gar,gar0,gar1,gar2} :
\begin{equation}
\rho_*(x)=\frac{2}{\pi}\frac{1}{(1+x^2)^2}.\label{l8}
\end{equation}
The transition rate \eqref{l2} $w_\phi(z+x|x)$ for any $\mu \in (0,2)$  takes the form
\begin{equation}
w_\phi(z+x|x)=\frac{C_\mu}{|z|^{1+\mu}}\frac{1+x^2}{1+(z+x)^2}.\label{l9}
\end{equation}
Similar to the  previous Gaussian case, the simulations can be speed up by analytical evaluation of some integrals.
 Such   acceleration of numerical routines  permits to handle (in the same timescale) trajectories for much longer running
  times ($t\sim 400$)  than in the previous harmonic  case ($t\sim 10$).
Each $\mu $ -driver case  ($\mu=0.5, 1, 1.5$) will be addressed separately.
\begin{description}
\item [Case of] $\mu=1/2$.   For $z>0$ we  need to evaluate
\end{description}
\begin{equation}
f_{1/2}(x,z)=\int\frac{1}{z^{3/2}}\frac{1+x^2}{1+(z+x)^2}dz.\label{l10}
\end{equation}
As
\begin{equation}
\frac{1}{z^{3/2}}\frac{1+x^2}{1+(z+x)^2}=\frac{1}{z^{3/2}}-\frac{2x+z}{\sqrt{z}\left[1+(x+z)^2\right]},\label{l11}
\end{equation}
 $f_{1/2}(x,z)$ can be written as
\begin{equation}
f_{1/2}(x,z)=-2z^{-1/2}-\int\frac{2x+z}{\sqrt{z}(1+(x+z)^2)}dz.\label{l12}
\end{equation}
For small  $z$ the dominant contribution comes from the first term so that the second term can
 efficiently be evaluated numerically. We note here that although the integration in the second term of
 \eqref{l12} can be performed analytically, the result appears to be quite cumbersome. Therefore, for our purposes it is more profitable
  to integrate this term numerically. For  $z<0$  the corresponding integral can be expressed as $- C_\mu   f_{1/2}(-x,-z)$.
\begin{description}
\item [Case of] $\mu=1$. Here, we need   to  evaluate the  integral
\end{description}
\begin{widetext}
\begin{equation}
f_1(x,z)=\int\frac{1+x^2}{1+(z+x)^2}\cdot\frac{dz}{z^2}=\frac{x}{1+x^2}\ln\left(1+(z+x)^2\right)+
\frac{x^2-1}{x^2+1}\arctan(z+x)-\frac{2x}{1+x^2}\ln|z|-\frac{1}{z}.\label{l16}
\end{equation}
\end{widetext}
We pay attention to the fact  that  the integrand is rational,  hence the integral  can be evaluated  analytically in a simple form.
This permits to use the analytical answer \eqref{l16} in our simulations without the  need to divide the integration range
into "small" and "large" $z$ domains.
\begin{description}
\item[Case of]  $\mu=3/2$. In this case, we have for $z>0$
\end{description}
\begin{equation}
f_{3/2}(x,z)=\int\frac{1}{z^{5/2}}\frac{1+x^2}{1+(z+x)^2}dz. \label{l17}
\end{equation}
Since
\begin{equation}
\frac{1}{z^{5/2}}\frac{1+x^2}{1+(z+x)^2}=\frac{1}{z^{3/2}}\left(\frac{1}{z}-\frac{2x+z}{1+(x+z)^2}\right),
\label{l17a}
\end{equation}
there holds
\begin{eqnarray}
f_{3/2}(x,z)&=&-\frac{2}{3z^{3/2}}+\frac{4x}{(1+x^2)z^{1/2}}+\nonumber \\
&&\int\frac{2x z+3x^2-1}{(1+x^2)(1+(x+z)^2)z^{1/2}}dz. \label{l17b}
\end{eqnarray}
Again, the third term in \eqref{l17b} can efficiently be evaluated numerically.
For  $z<0$   we encounter   $-f_{3/2}(-x,-z)$.
\begin{figure*}
\centerline{\includegraphics[width=0.7\columnwidth]{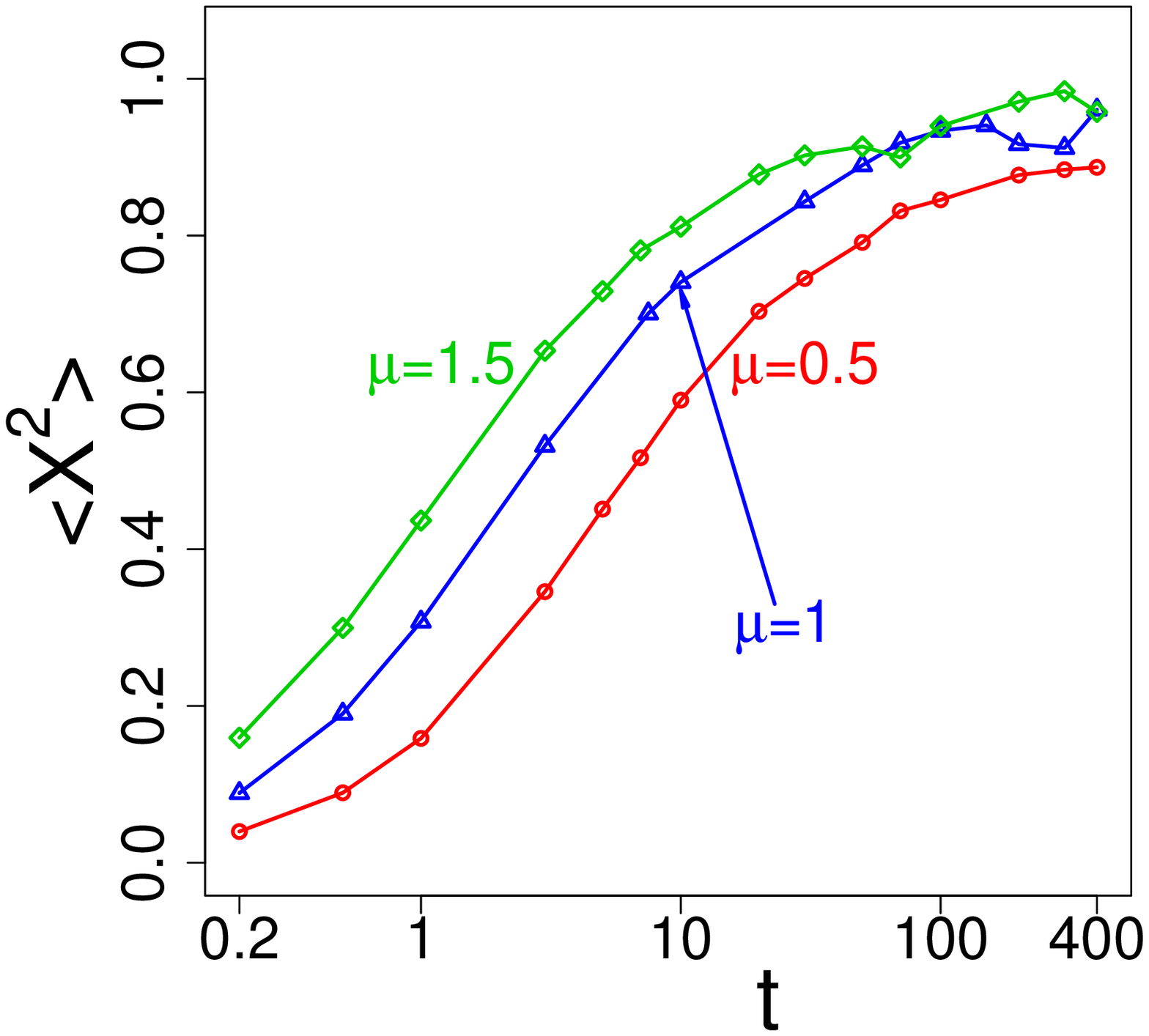} \quad
\includegraphics[width=0.7\columnwidth]{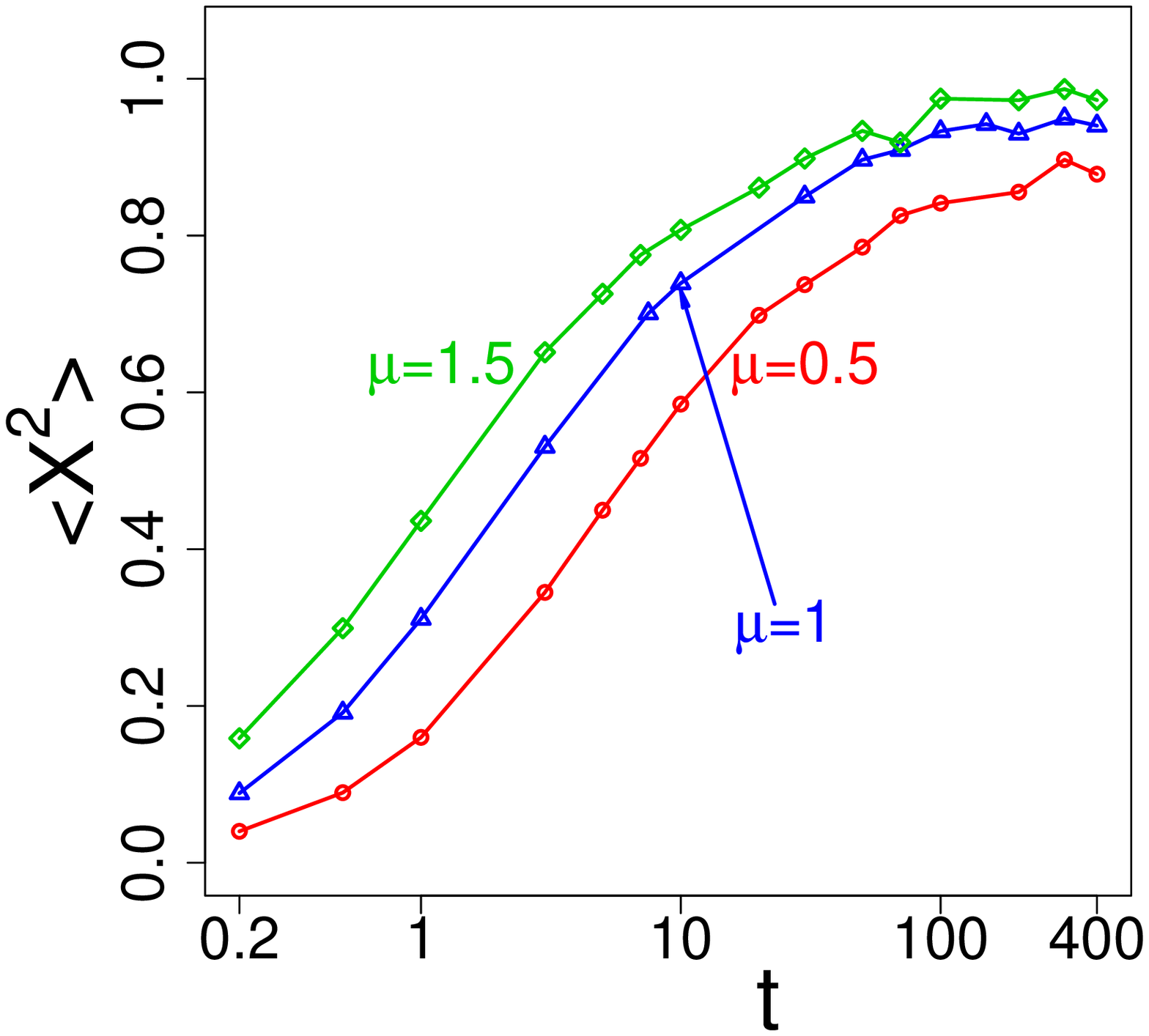}}
\centerline{\includegraphics[width=0.7\columnwidth]{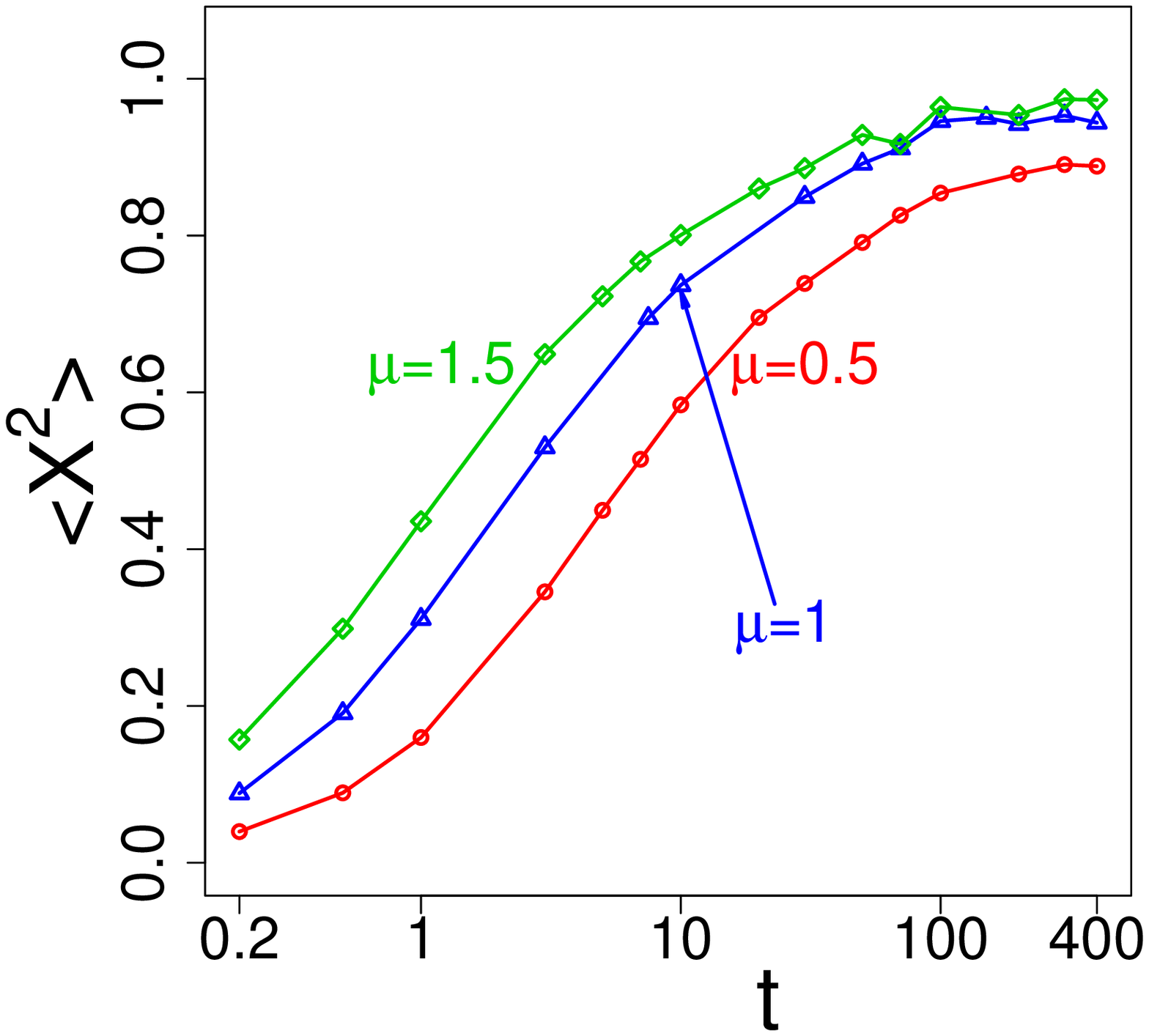} \quad
\includegraphics[width=0.7\columnwidth]{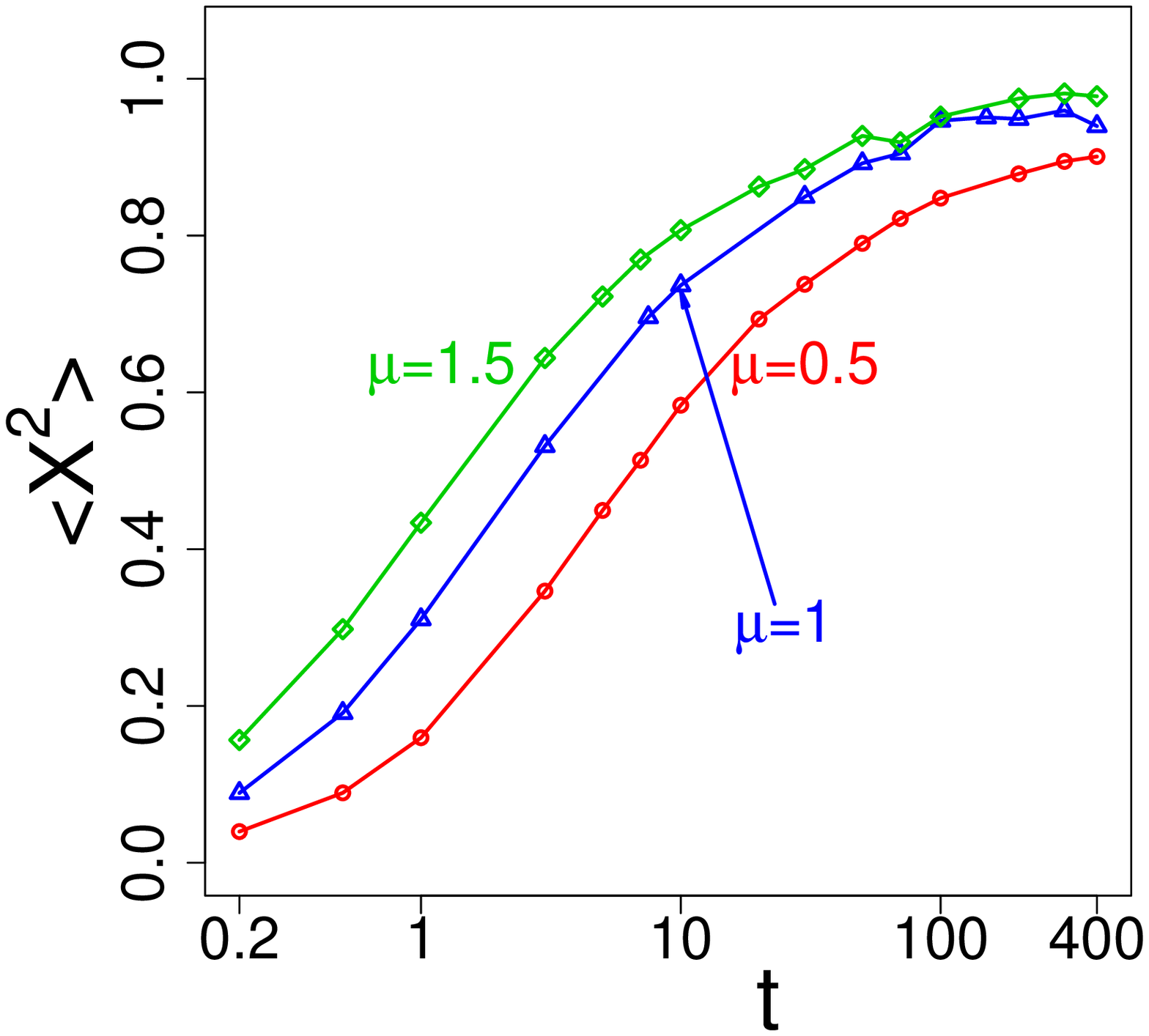}}
\caption{Quadratic Cauchy target: Time evolution of the pdf $\rho (x,t)$ second moment for 50 000 (upper left panel), 100 000 (upper right panel), 150 000 (lower left panel) and 200 000 (lower right panel) trajectories.}
\end{figure*}

\begin{figure*}
\centerline{
\includegraphics[width=0.7\columnwidth]{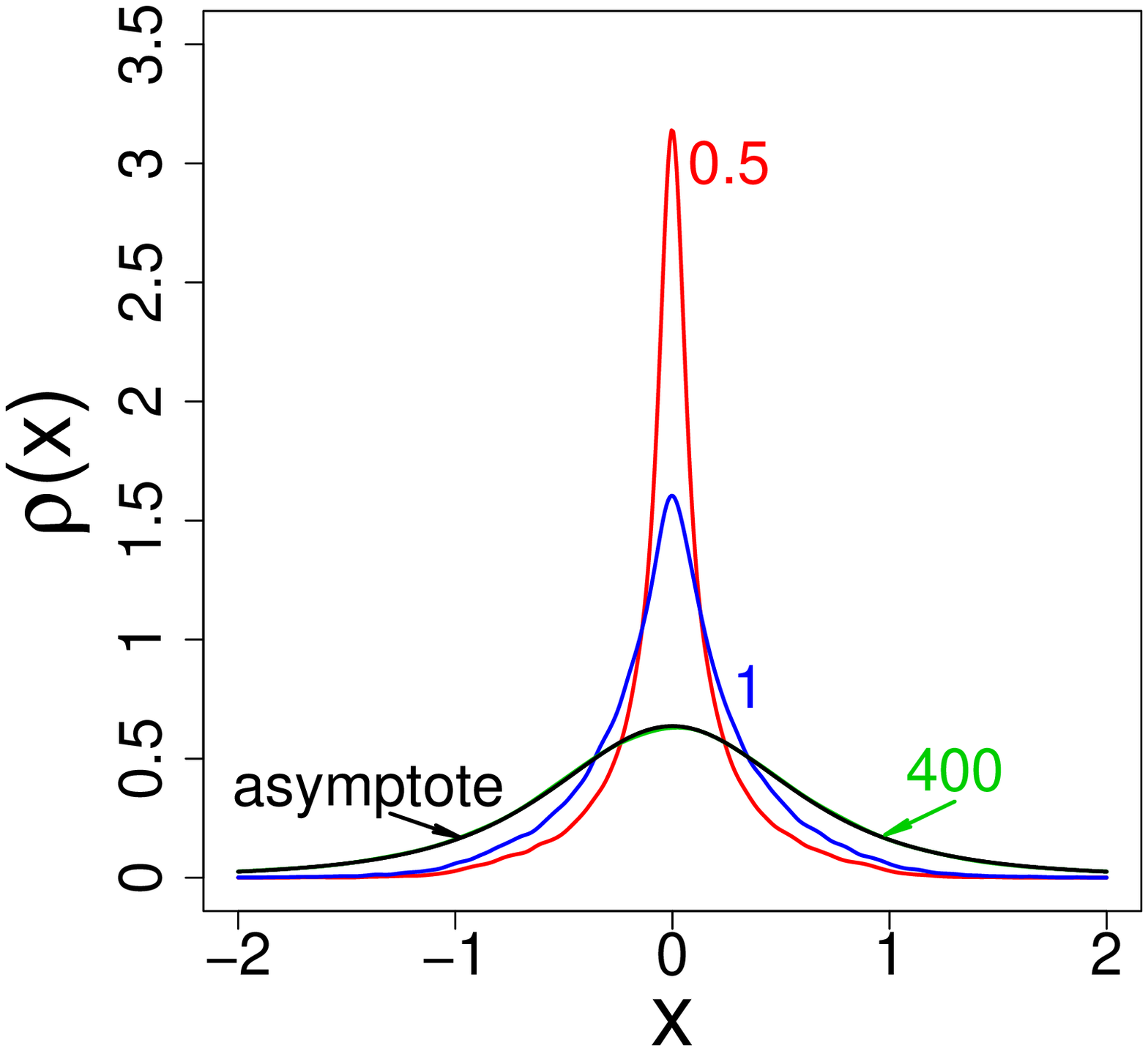}
\includegraphics[width=0.7\columnwidth]{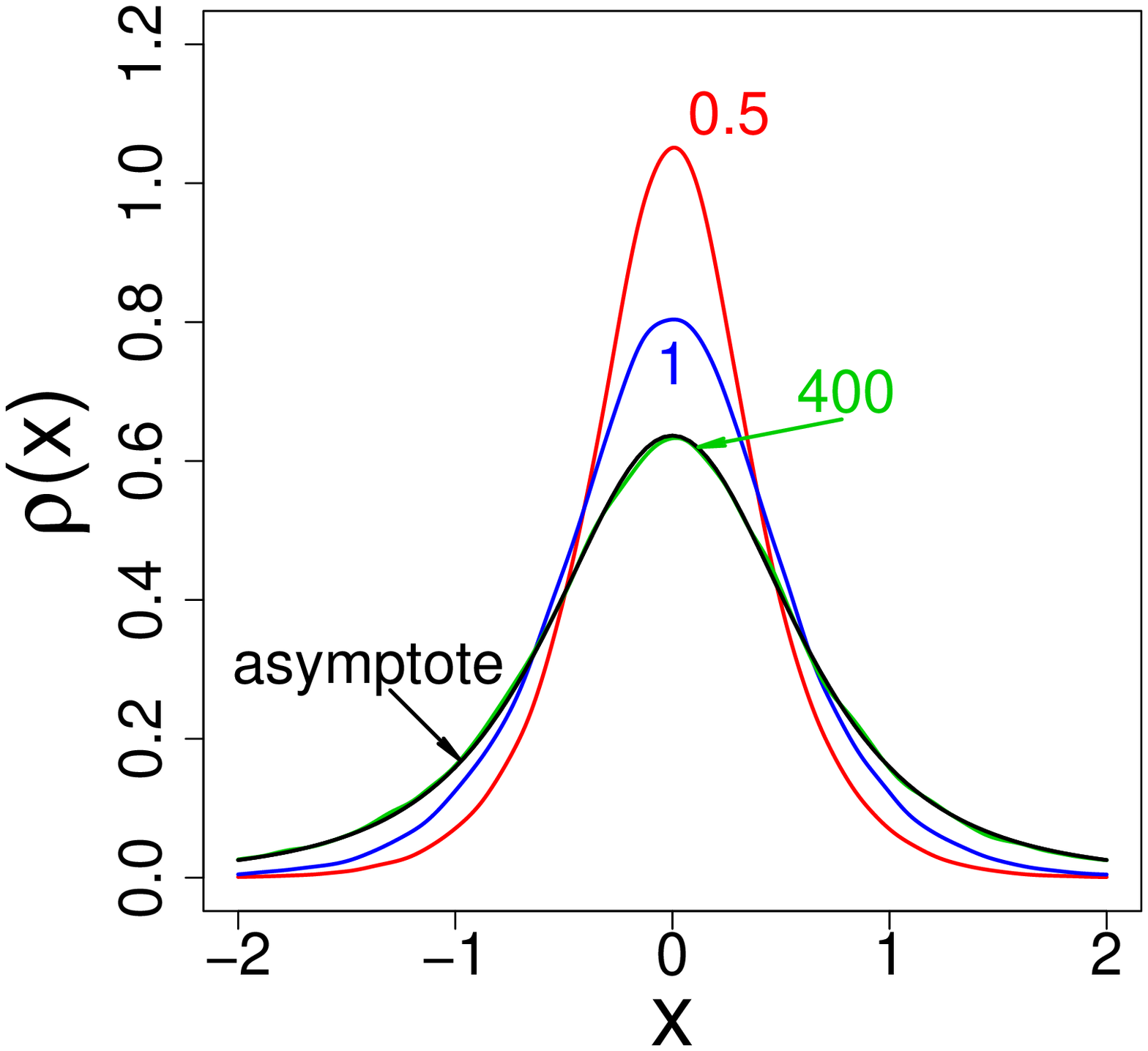}
\includegraphics[width=0.7\columnwidth]{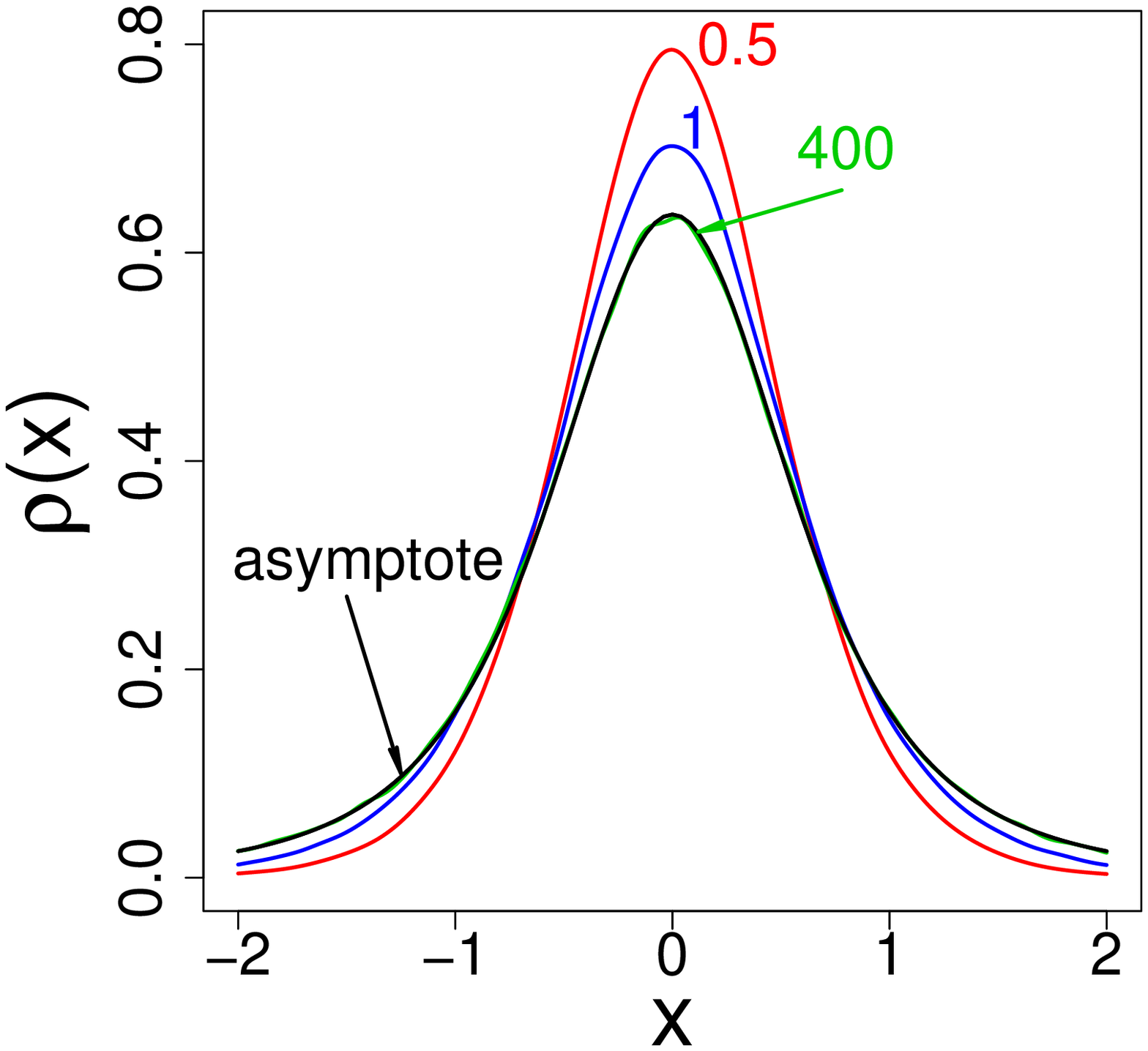}}
\caption{Quadratic Cauchy target: Time evolution of  $\rho (x,t)$ inferred from  $200 000$ trajectories for
$\mu=0.5$ (left panel), $\mu=1$ (middle panel) and $\mu=1.5$ (right panel).  All trajectories are started from $x=0$.
Note scale differences on vertical axes.}
\label{rys5}
\end{figure*}

Simulation results are displayed in Figs. 4 and 5. If we compare Fig. 4 with Fig. 1 we see the existence of small oscillations
in the asymptotic regime about the value $1/2$.  Those  from Fig.1 were relatively small and  were  quickly  smoothed out with the growth
 of the number of trajectories used to extract statistical data. In Fig. 4  the oscillations are more noticeable  and persist even for
  $200 000$ trajectories and more. This is related to much slower decay of transition rates \eqref{l9}
 (determined by slow-decaying asymptotic pdf \eqref{l8}) as compared to those for Gaussian case \eqref{l7}.

The second moment of the present  $\rho_*(x)$, (\ref{l8}), equals 1 and the convergence towards this value  is clearly seen in Fig. 3.
 This convergence is much slower than in the Gaussian  (harmonic confinement) case  which is not a surprise:  (\ref{l7}) and  (\ref{l9}))
   indicate that the present  rate of convergence should be logarithmically slower. Fig. 5,  quite alike Fig. 2, convincingly
  demonstrates a convergence  of $\rho (x,t)$ to the asymptotic $\rho _*(x)$.
  For definitely large  times around $t=400$,   $\rho (x,t)$  and $\rho _*(x)$  become   practically indistinguishable.
  Similarly  to the Gaussian case, the rate of convergence becomes larger  with the growth of   $\mu \in (0,2)$.

\begin{figure*}
\centerline{
\includegraphics[width=0.7\columnwidth]{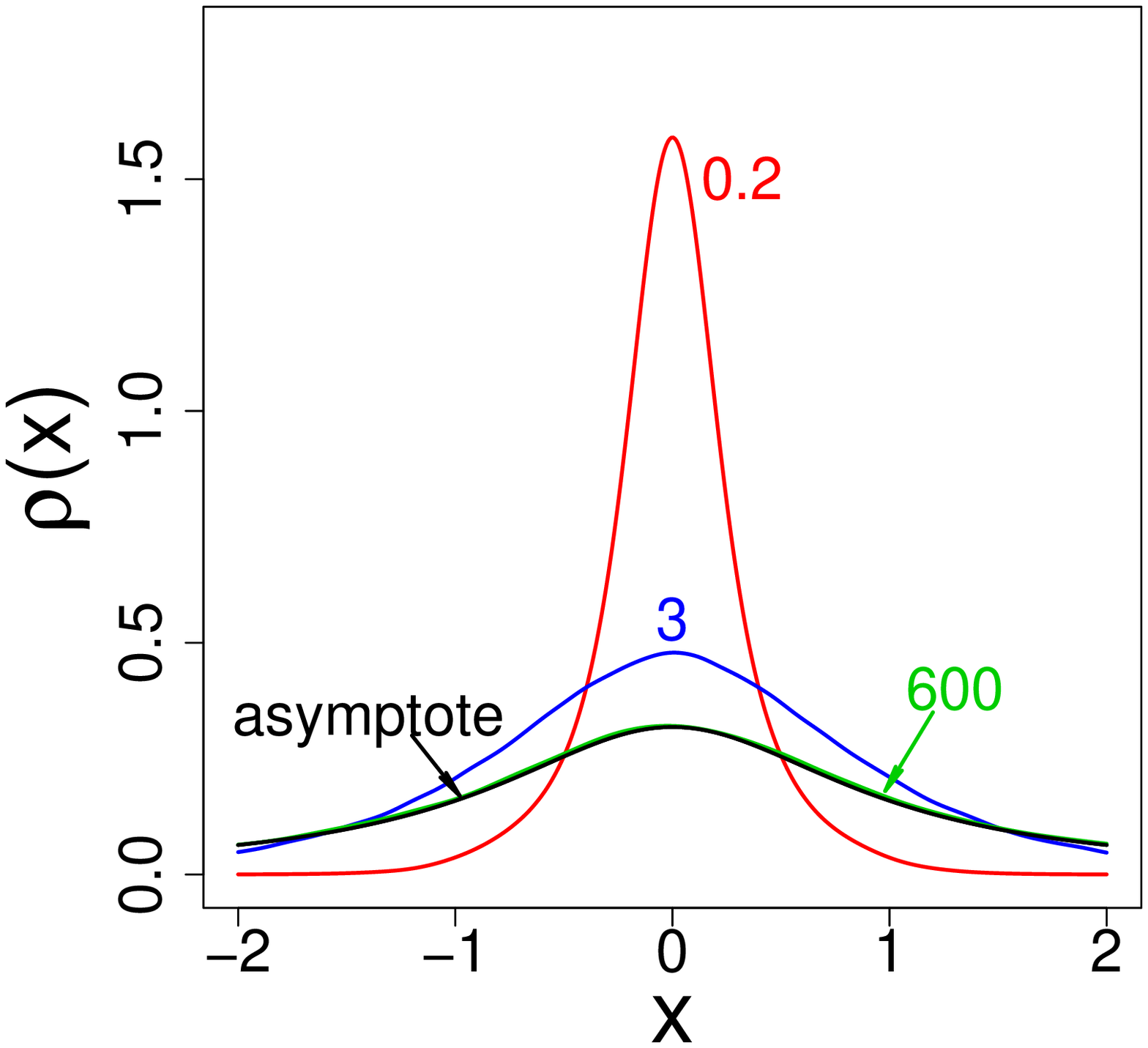}
\includegraphics[width=0.7\columnwidth]{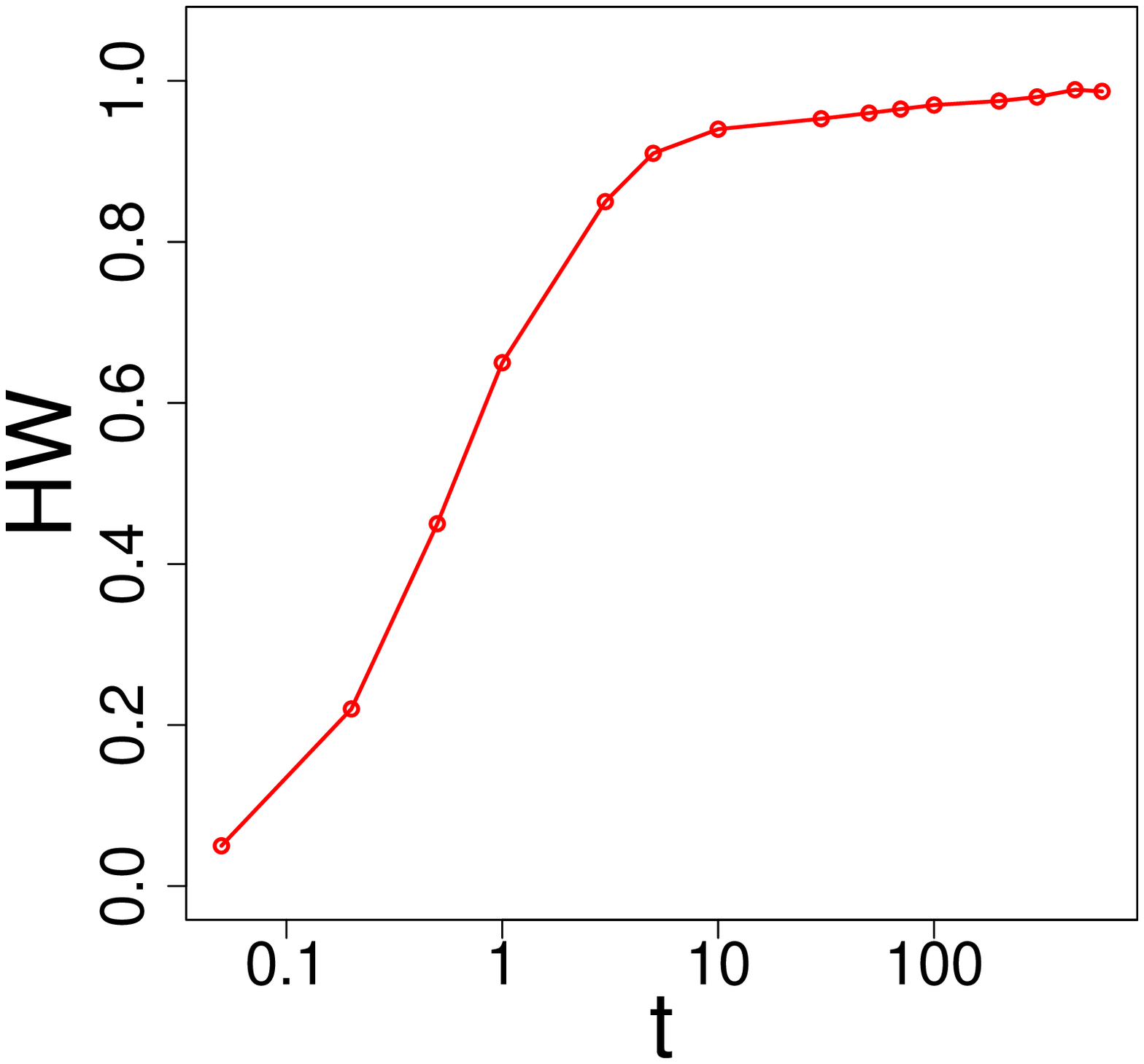}
\includegraphics[width=0.7\columnwidth]{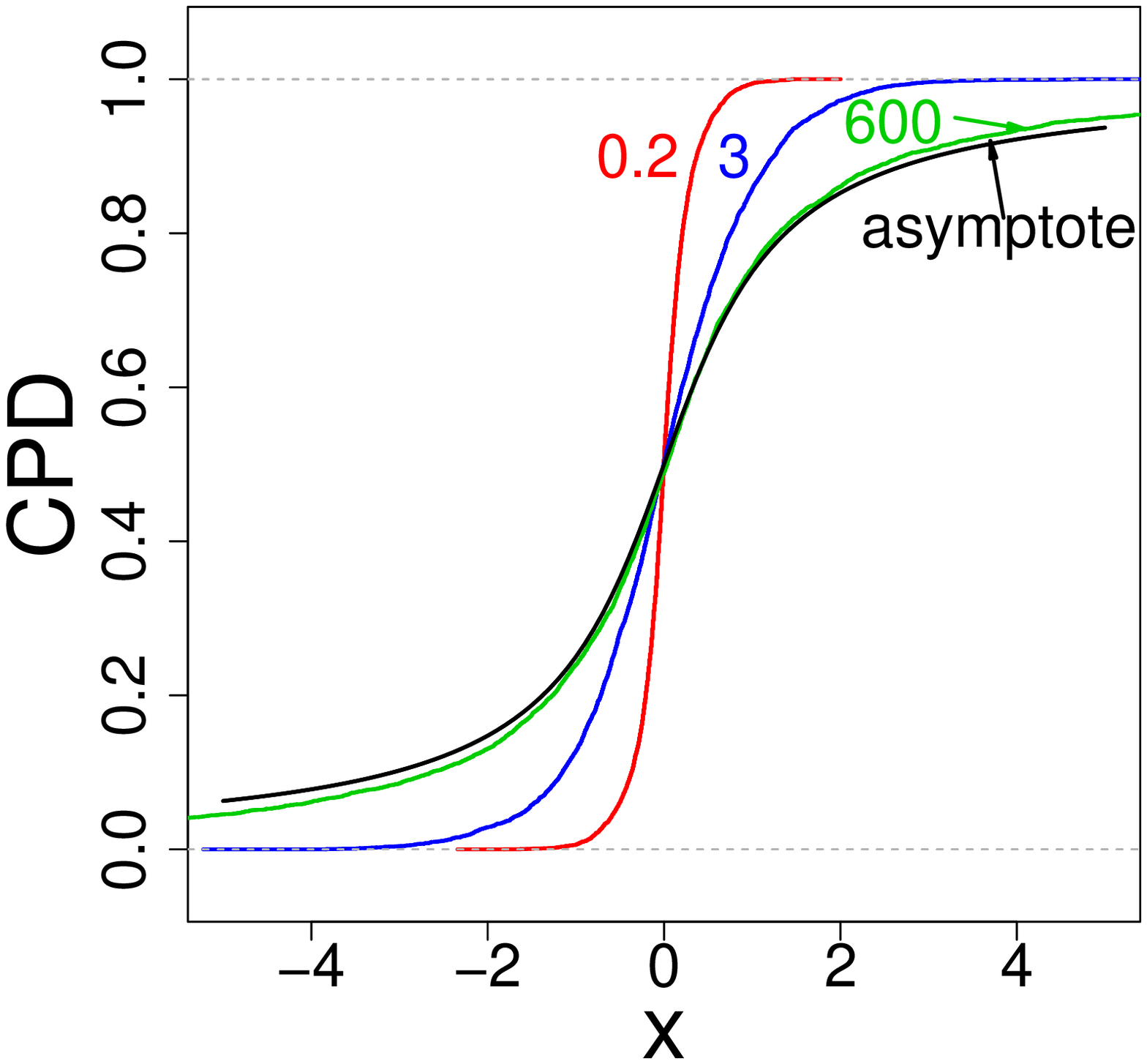}}
\caption{Cauchy target: Time evolution of pdf $\rho (x,t)$ (left panel), half-width (HW) of  $\rho (x,t)$ (middle panel).
 Right panel reports the cumulative  probability distributions (CPD) for different time instants.
 Here $\mu =1$ and all data are inferred from $200 000$ trajectories, starting from $x=0$.}
\label{rys6}
\end{figure*}

\subsubsection{Cauchy target}
Now we consider  an asymptotic pdf  of the form :
\begin{equation}
\rho_*(x)=\frac{1}{\pi}\frac{1}{1+x^2}.\label{l20}
\end{equation}
In this case, the transition rate from $x$  to  $x+z$  reads
\begin{equation}
w_\phi(z+x|x)=\frac{C_\mu}{|z|^{1+\mu}}\sqrt{\frac{1+x^2}{1+(z+x)^2}}.\label{l21}
\end{equation}
We consider  Cauchy driver corresponding to $\mu=1$. The transition rate integral can be evaluated analytically. For   $z>0$ we have
\begin{widetext}
\begin{equation}
f(x,z)=\int\sqrt{\frac{1+x^2}{1+(z+x)^2}}\cdot\frac{dz}{z^2}=\frac{-\sqrt{1+x^2}\sqrt{1+(x+z)^2}+xz\ln\left((1+x^2+xz+\sqrt{1+x^2}\sqrt{1+(x+z)^2})/z\right)}{(1+x^2)z}. \label{l21}
\end{equation}
\end{widetext}
For  $z<0$  the outcome is $-f(-x,-z)$.

In Fig. 6, we report the time evolution of the statistically inferred  $\rho (x,t)$,  its half-width
(as second moment does not exist for asymptotic pdf \eqref{l20}) and simulated cumulative  probability  distributions (CPD)
for different time instants. An approach to the asymptotic pdf  (\ref{l20}) is clearly seen, together with a convergence of
a half-width to its asymptotic value 1.  The same convergence pattern is observed for CPD which approaches the
 asymptotic function $F(x)=\frac{1}{2}+\frac{\arctan x}{\pi}$.

\textbf{Comment 3}:  Displayed empirical (numerically retrieved) curves  in Fig. 6  are hampered by certain errors.
 The figures have been read from a histogram of randomly sampled data.  Its  partitioning  into subintervals is  a source of inaccuracies. In case of a small number of  intervals, the read-out
 error would be large,  with a size of  about half-interval length.
A finer partitioning (large number of small subintervals)  would still produce an error which is
close  to the half-maximum of the curve.
The error bound would be smaller  or  equal to the half-length of  subintervals  corresponding to roughly the same histogram values.
 One more inaccuracy source in the finer partition case comes from the maximum read-out imprecision. Namely, we can have a conspicuous peak, whose close  vicinity displays much (half or less) smaller    values.    Therefore the partitioning finesse must be slightly optimized.

\subsection{Locally  periodic confinement}

To set firm grounds for future research it is instructive to study our model for more complicated forms of confining potentials.
In view of their physical relevance,  it is appealing to address an issue of confining (trapping) environments with a periodic spatial structure.
Here, we encounter a major difficulty  with a  $L^1(R)$ integrability of the Boltzmann-type weighting function $\exp(-\Phi )$.
Periodicity and integrability can here be reconciled either on compact  sets or by means of locally periodic potentials  that take a definite confining form (harmonic or polynomial) for larger values of $x\in R$.
Let us consider the following asymptotic pdf
\begin{equation}
\rho_*(x)=\left\{
            \begin{array}{ll}
              \frac{1}{C} e^{-\sin^2(2\pi x)}, & \hbox{$|x|\leqslant 2$;} \\
              \frac{1}{C} e^{-(x^2-4)}, & \hbox{$|x|>2$,}
            \end{array}
          \right.
\end{equation}
where $C=3.032818$  is a normalization constant.
The transition rate from  $x$ to  $x+z$  reads
\begin{equation}
w_\phi(z+x|x)=\frac{C_\mu}{|z|^{1+\mu}}\exp{\left[(\phi(x)-\phi(x+z))/2\right]},
\end{equation}
where the  potential  $\phi$  has the form
\begin{equation}
\phi(x)=\left\{
          \begin{array}{ll}
            \sin^2(2\pi x), & \hbox{$|x|\leqslant 2$;} \\
            (x^2-4), & \hbox{$|x|>2$.}
          \end{array}
        \right.
\end{equation}
We consider  $\mu=1$.  To optimize the simulation, here we use the same trick of isolating of "most dangerous" small $z$
terms in the integrals involved in the Gillespie algorithm. For small  $z$
we expand  the  term  $\exp[(\sin^2(2\pi x)-\sin^2(2\pi(x+z)))/2]$ in Taylor series.  We choose
$\varepsilon_{12}=0.05$. In the vicinity of $|x|=2$ due  attention  must be paid to the proper power series truncation,
 to correctly choose the intervals where integration should be performed numerically.
 For example at  $x\in(1.95,2)$ we   have
 \begin{widetext}
\begin{eqnarray}
&&\frac{1}{C_1}\int\limits_{\varepsilon_1}^{\varepsilon_{12}}w_\phi(z+x|x)dz
=\int\limits_{\varepsilon_1}^{2-x}\exp{\left[\frac{\sin^2(2\pi x)-\sin^2(2\pi(x+z))}{2}\right]}\frac{dz}{z^2}+\int\limits_{2-x}^{\varepsilon_{12}}\exp{\left[\frac{\sin^2(2\pi x)-(x+z)^2+4}{2}\right]}\frac{dz}{z^2} \nonumber \\
&&=\int\limits_{\varepsilon_1}^{2-x}\exp{\left[\frac{\sin^2(2\pi x)-\sin^2(2\pi(x+z))}{2}\right]}\frac{dz}{z^2}+\exp{\left[\frac{\sin^2(2\pi x)-x^2+4}{2}\right]}\int\limits_{2-x}^{\varepsilon_{12}}\exp{\left[\frac{x^2-(x+z)^2}{2}\right]}\frac{dz}{z^2}.
\end{eqnarray}
 \end{widetext}
The numerators of integrand fractions have been expanded into Taylor series and (safely) truncated at the quadratic terms.

\begin{figure}[h]
\begin{center}
\includegraphics[width=0.7\columnwidth]{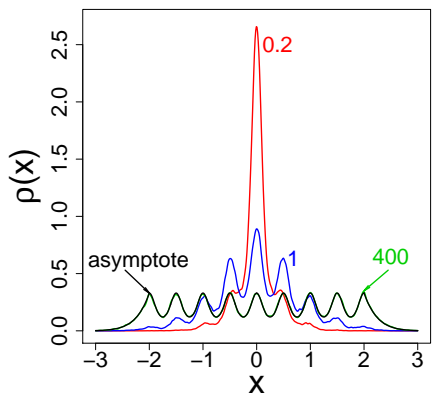}
\end{center}
\caption{Time evolution of $\rho (x,t)$ inferred from  $200 000$  trajectories at $\mu =1$. The  data for $100 000$ and $300 000$ trajectories
(not displayed) do not show qualitative differences.}
\end{figure}

Time evolution of the inferred  pdf $\rho (x,t)$ is reported in Fig. 7. All sample trajectories were started form $x=0$ which
 corresponds to the $\delta (x) $-type initial distribution. The probability density spreads out
with time in conformity with the trapping (confining) properties of the locally periodic enclosure
(environment or "potential landscape"). For large running times t=400 the trajectory statistics  produces data
 that are indistinguishable from those for the asymptotic pdf. We have checked that beginning  from about  100 000 trajectories,
  further accumulation of the trajectories number like e.g. 200 000 (displayed) and 300 000  (not displayed)
  for the data statistics is inessential. In such cases the curves are almost the same,  we  merely improve a fidelity of the statistics.

\section{Conclusions}

If a random process does not admit the description in terms of a stochastic differential equation  (e.g. Langevin modeling), its direct numerical simulation
becomes  impossible  by means of  existing popular  algorithms. In the present paper, for the first time in the literature, we propose  a working method
to generate stochastic trajectories (sample paths) of a random jump-type process without resorting to any explicit (or numerical) solution of a stochastic
 differential equation. To this end we have  modified    the Gillespie algorithm \cite{gillespie,gillespie1},
normally devised for sample paths generation   if the   transition rates  refer to  a finite  number of states of  a system.

  The essence of our   modification is that we take into account the continuum of possible transition rates, thereby changing the finite  sums
  in the original Gillespie algorithm into  integrals.  The corresponding  procedures for  stochastic trajectories generation has  been
  changed accordingly. In other words, here we "extract" the background sample paths of a jump process, whose pdf obeys the  transport equation
  (generalized Fokker-Planck dynamics)  \eqref{l1},  \eqref{l2}. We emphasize once more  here, that  we have focused  on those background jump-type
  processes that  cannot be  modeled  by any stochastic differential equation of the  Langevin  type.

Although heavy-tailed  L\'{e}vy stable drivers were involved in the present considerations,
 we have clearly confirmed that an enormous variety of stationary  target  distributions is  dynamically accessible
   in each particular $\mu \in (0,2)$  case. That comprises not only a standard Gaussian pdf,  casually  discussed in relation
   to the Brownian  motion (e.g. the  Wiener process). Among heavy-tailed distributions, we have paid attention to the  Cauchy pdf
   which can stand for  an asymptotic target for  any $\mu \neq 1$ driver,
    provided a steering environment is properly devised. In turn, the Cauchy driver  in a proper  environment
     may lead to an asymptotic pdf with a finite (in fact arbitrarily large) number of moments,  the Gaussian case being  included (\cite{gar}).

An example of the locally periodic environment has been considered as a toy model for more  realistic  physical systems.
Our major hunch are  strongly  inhomogeneous  "potential landscapes", \cite{gar2},
 being  sufficiently smooth to avoid a direct  reference to random potentials, \cite{geisel}.  Even if  various mean field data
 are available in such (experimentally realizable) systems, it is of interest to have some knowledge about the microscopic \
 dynamics (random paths) for the system under consideration.   The detailed analysis of sample path data (ergodicity, mixing or
  lack of those properties) deserve a separate analysis.

We mention  possible  generalizations of our method to  the  Brownian motor concept  (see, e.g., Ref. \cite{browmot} for recent review)
 to include a  non-Gaussian jumping component. In those systems it  is   the  properly tailored periodic "potential landscape"
  which enforces a   conversion  of  a homogenous   stochastic process (Brownian motion for reference)
  into the directed motion of  particles at nanometer scales.
  That  is closely related to the problem of so-called
 sorting in periodic potentials \cite{presort}. Other problem to be addressed  concerns  ultracold atoms in optical lattices subject to
 random potentials \cite{coldopt}, which might  promising not only from a purely scientific point of view, but also with prospects for many
 technological applications.
 We note  that the  theoretical description of the  above   mentioned topics  relies  essentially on the
  Langevin-like equation  input.

   Our approach  offers an immediate generalization for generalizations, where
 systems with non-Langevin response to external potentials may come  into  consideration, along with  more traditional ones.
 What we actually  need to implement our version of  Gillespie's algorithm   is the knowledge of
  jump  transition rates of those random  systems only.

A  preliminary work (in progress) shows that an extension of our algorithm to higher dimensions is operational.
In particular, the planar case is worth exploration, possibly with more  complex  "potential landscapes".
While departing from  final comments of  Ref. \cite{gar2} we  expect that the presented methodology can be  effectively
 adopted to construct optimal random search routines, see in this connection \cite{snider}.

\textbf{Acknowledgement}:  P. G. would like to thank   Igor M. Sokolov for indicating a possible  relevance  of the Gillespie's algorithm out of  its  original  chemical kinetics  context.  All numerical  simulations were completed by means of the facilities of
the  Platon - Science Services Platform of the Polish Pionier Network.

\end{document}